\documentclass[a4paper,12pt,numbers,sort&compress]{article}
\pdfoutput=1
\usepackage{amsfonts,amsmath,amssymb}
\usepackage[paper=letterpaper,margin=1.0in]{geometry}
\usepackage{graphicx}
\usepackage{cite}
\usepackage{hyperref}

\def\Tiny{\fontsize{6pt}{6pt}\selectfont}

\def\mP{\mathbb{P}}

\newcommand{\be}{\begin{equation}}
\newcommand{\ee}{\end{equation}}
\newcommand{\bea}{\begin{eqnarray}\displaystyle}
\newcommand{\eea}{\end{eqnarray}}
\newcommand{\nnm}{\nonumber}

\newcommand{\ba}{\begin{array}}
\newcommand{\ea}{\end{array}}
\newcommand{\ben}{\begin{enumerate}}
\newcommand{\een}{\end{enumerate}}
\newcommand{\bi}{\begin{itemize}}
\newcommand{\ei}{\end{itemize}}
\newcommand{\bc}{\begin{center}}
\newcommand{\ec}{\end{center}}
\newcommand{\bfig}{\begin{figure}}
\newcommand{\efig}{\end{figure}}
\newcommand{\bq}{\begin{quotation}}
\newcommand{\eq}{\end{quotation}}
\newcommand{\bt}{\begin{table}}
\newcommand{\et}{\end{table}}
\newcommand{\btab}{\begin{tabular}}
\newcommand{\etab}{\end{tabular}}
\newcommand{\bmi}{\begin{minipage}}
\newcommand{\emi}{\end{minipage}}
\newcommand{\bs}{\begin{slide}}
\newcommand{\es}{\end{slide}}
\newcommand{\nn}{\nonumber}
\newcommand{\eref}[1]{(\ref{#1})}
\newcommand{\cN}{{\cal N}}

\newcommand{\bmQ}{\overline{\mathbb{Q}}}

\newcommand{\mQ}{ \mathbb{Q} }

\newcommand{\IC}{{\mathbb C}}
\newcommand{\IF}{{\mathbb F}}

\newcommand{\IP}{{\mathbb P}}
\newcommand{\IQ}{{\mathbb Q}}
\newcommand{\IZ}{{\mathbb Z}}

\newcommand{\tr}{ {\rm Tr} }

\newcommand{\cM}{ \mathcal{M} }

\newcommand{\cD}{\mathcal{D}}

\newcommand{\pa}{\partial}

\newcommand{\setall}{\setcounter{equation}{0}}
\renewcommand{\thefootnote}{\fnsymbol{footnote}}

\newcommand{\comment}[1]{}

\begin{document}
\rightline{IMPERIAL/TP/11/AH/06}
\rightline{QMUL-PH-11-05}

\vskip 1cm

\centerline{{\LARGE \bf  Invariants of Toric Seiberg Duality }} 
\medskip

\vspace{.4cm}

\centerline{
{\large Amihay Hanany}$^1$,
{\large Yang-Hui He}$^2$,
{\large Vishnu Jejjala}$^3$,
{\large Jurgis Pasukonis}$^3$,
}
\vspace{.4cm}
\centerline{
{\large Sanjaye Ramgoolam}$^3$,
{\large and}
{\large Diego Rodriguez-Gomez}$^4$
\footnote{
a.hanany@imperial.ac.uk,
yang-hui.he.1@city.ac.uk,
v.jejjala@qmul.ac.uk,
j.pasukonis@qmul.ac.uk,
s.ramgoolam@qmul.ac.uk,
drodrigu@physics.technion.ac.il}
}

\vspace*{3.0ex}

\begin{center}
{${}^{1}$ Theoretical Physics Group, The Blackett Laboratory,\\
Imperial College, Prince Consort Road, London SW7 2AZ, UK\\
}
\vspace*{1.5ex}
{${}^{2}$ Department of Mathematics, City University, London,\\
Northampton Square, London EC1V 0HB, UK;\\
School of Physics, NanKai University, Tianjin, 300071, P.R.~China;\\
Merton College, University of Oxford, OX14JD, UK\\
}
\vspace*{1.5ex}
{${}^{3}$ Department of Physics, Queen Mary, University of London,\\
Mile End Road, London E1 4NS, UK\\
}
\vspace*{1.5ex}
{${}^{4}$ Department of Physics, Technion, Haifa, 3200, Israel;\\}
{Department of Mathematics and Physics,\\
University of Haifa at Oranim, Tivon, 36006, Israel\\
}
\end{center}

\vspace*{4.0ex}
\centerline{\textbf{Abstract}} \bigskip

Three-branes at a given toric Calabi--Yau singularity lead to different phases of the conformal field theory related by toric (Seiberg) duality.
Using the dimer model/brane tiling description in terms of bipartite graphs on a torus, we find a new invariant under Seiberg duality, namely the Klein $j$-invariant of the complex structure parameter in the distinguished 
isoradial embedding of the dimer, determined by the physical $R$-charges. 
Additional  number theoretic invariants are described 
in terms of the algebraic number field of the  $R$-charges. 
We also  give a new compact description of the  $a$-maximization procedure 
by introducing a generalized incidence matrix.

\newpage

\tableofcontents

\newpage

\renewcommand{\thefootnote}{\arabic{footnote}}
\setcounter{footnote}{0}

\section{Introduction}\setall

Thanks to the seminal work of N.~Seiberg, strong coupling/weak coupling duality in non-Abelian $\mathcal{N}=1$ super-Yang--Mills (SYM) theories is well-established~\cite{Seiberg:1994pq}.\footnote{
We should note that no strict proof has been given for generic cases.
Nevertheless, there is virtually no doubt on the validity of the duality.}
This duality has been used and tested extensively in various environments, and it is now part of the standard toolkit of the (supersymmetric) field theorist.

Seiberg duality is a purely field theoretic statement.
Nevertheless, it is often interesting to examine it from a string theoretic perspective.
A ``proof'' of the duality may be established by means of operations in a brane configuration~\cite{Elitzur:1997fh}.
In this paper we will mostly be interested in gauge theories arising from D$3$-branes probing a toric Calabi--Yau threefold (CY$_3$) singularity.
As first noted in~\cite{Feng:2000mi, Feng:2001bn, Beasley:2001zp, Feng:2001xr}, within this context Seiberg duality is the field theory manifestation of \textit{toric duality}, wherein, different, but equivalent, descriptions of the same CY$_3$ are related by Seiberg duality of the associated field theories.
By now a heavy mathematical technology has been developed to understand the class of theories under discussion in which different nuances of Seiberg duality have been appreciated (see \textit{e.g.},~\cite{Cachazo:2001sg, Feng:2002kk, Wijnholt:2002qz, Herzog:2004qw, Berenstein:2002fi, Braun:2002sb, Murthy:2006xt, Forcella:2008bb, Forcella:2008ng}).

D$3$-branes probing conical CY$_3$ geometries are of special interest, as they give rise to examples of the AdS$_5$/CFT$_4$ correspondence~\cite{Maldacena:1997re, Gubser:1998bc, Witten:1998qj} with $\mathcal{N}=1$ supersymmetry.
These dualities have been particularly studied when the CY$_3$ is toric.
In fact, infinite series of explicit examples where both metric and field theory are known have been presented in~\cite{Benvenuti:2004dy, Benvenuti:2005ja, Franco:2005sm, Butti:2005sw}.
Further progress was made in~\cite{Franco:2005rj} where it was shown that these field theories ---
which flow to non-trivial IR superconformal fixed points ---
dual to D$3$-branes probing toric CY$_3$ geometries, can be neatly encoded in a bipartite graph drawn on a torus;
this graph is called a \textit{dimer}~\cite{Hanany:2005ve} or \textit{brane tiling}~\cite{Franco:2005rj}.
In~\cite{Feng:2005gw}, it was shown how the dimer is related, through mirror symmetry, to a system of intersecting D$6$-branes.
This mirror Type IIA configuration is as well related by T-duality to a five brane system from which the dimer can be read.
We refer to~\cite{Kennaway:2007tq, Yamazaki:2008bt} for thorough reviews.

A further refinement on the dimer was introduced in~\cite{Hanany:2005ss}, where it was shown how the $R$-charges can be encoded in certain angles in the dimer once it is drawn in a particular, \textbf{isoradial} manner.
In fact, isoradial embeddings for dimers have been introduced in the mathematical literature (for a review see \textit{e.g.},~\cite{Kenyon-review}).
When the angles in the isoradial dimer are associated to the 
physical $R$-charges (which maximize the trial $R$-charges 
in the $a$-maximization procedure), a distinguished complex structure (shape) 
of the torus is determined.  
This complex structure is denoted as $\tau_R$ and has been 
studied in ~\cite{Jejjala:2010vb, hhjprr}.

In~\cite{Jejjala:2010vb}, it was noted that the mathematical theory of \textit{dessins d'enfants} enables us to uniquely express bipartite graphs on a Riemann surface $\Sigma$ equivalently in terms of a holomorphic map $\beta~:~\Sigma \rightarrow \mathbb{P}^1$ with three branch points that can be chosen to lie at $\{0, 1, \infty\}$. The pull-back of the unique complex structure of 
the $\mP^1$ to $\Sigma$ determines a complex structure on $\Sigma$. 
For the case of a dimer drawn on a torus $\mathbb{T}^2$,
we have a  complex structure $\tau_B$ on an elliptic curve. 
In infinite families of examples, $\tau_B = \tau_R$ up to $SL(2,\mathbb{Z})$ equivalence, although we now know that the equality does not always hold~\cite{hhjprr}.\footnote{
Phase two of $L^{2,2,2}$ supplies a counterexample.
Nevertheless, there are many examples for which the equality does hold and it would be interesting to investigate how generic this situation is.
In fact, the counterexample found in~\cite{hhjprr} applies to the non-minimal phase of $L^{2,2,2}$ obtained by Seiberg dualizing the non-chiral Douglas--Moore orbifold of the conifold, which does satisfy $\tau_B=\tau_R$.
Thus, one might speculate that $\tau_B=\tau_R$ holds for minimal phases while the equality is somehow modified for non-minimal ones.
}

In this note we study Seiberg duality from the dimer model perspective, where it goes under the name ``\textbf{urban renewal}''~\cite{Propp:urbanrenewal,Franco:2005rj}.
Remarkably, the distinguished isoradial torus is the same for each of the different phases of a theory related by Seiberg duality.
That is to say, we can think of $\tau_R$ --- or more precisely the associated $j$-invariant $j(\tau_R)$ --- as an invariant under Seiberg duality.
In this note we will prove this by means of field theoretic methods.
It is natural to suspect, however, that this has a more profound geometrical meaning when regarded from a string theoretic perspective.
We will postpone this discussion to a forthcoming publication~\cite{SLAGS}.
Furthermore, motivated by the invariance of $j(\tau_R)$ among different Seiberg dual phases, we will introduce a new elegant repackaging of the $a$-maximization procedure~\cite{Intriligator:2003jj} for brane tilings~\cite{Butti:2005vn, Kato:2006vx}. This is done in terms of a generalized incidence matrix,   
which also gives a neat expression for further 
invariants under Seiberg duality.

The paper is structured as follows.
Section~\ref{SeibergAndDimer}, following a lightning review of dimer model technology for four-dimensional SCFTs, introduces the distinguished 
isoradial torus and its complex structure $\tau_R$.
By computing a large number of examples, we will motivate that $\tau_R$, or more precisely $j(\tau_R)$, is an invariant under Seiberg duality.
Section~\ref{sec:UrbanRen} provides a field theoretic proof of this statement.
Section~\ref{sec:taufromtoric} revisits $\tau_R$ from direct examination of the toric diagram.
Section~\ref{sec:five} gives a new description of the $a$-maximization of dimer models in terms of a \textit{generalized incidence matrix}.
Section~\ref{sec:six} considers further invariants of Seiberg duality.
Section~\ref{sec:conc} concludes.

\section{Dimer models and Seiberg duality}\setall
\label{SeibergAndDimer}

Dimer models were introduced in~\cite{Hanany:2005ve, Franco:2005rj, Feng:2005gw} as an extremely efficient way of packaging field theories arising from D$3$-branes at the tip of toric CY$_3$ cones over a five-dimensional Sasaki--Einstein base $\mathcal{B}$.
These theories flow in the IR to non-trivial SCFTs, which, by standard decoupling limit arguments~\cite{Maldacena:1997re, Gubser:1998bc, Witten:1998qj}, are dual to Type IIB supergravity on AdS$_5\times \mathcal{B}$.
We refer the reader to~\cite{Yamazaki:2008bt} for an extensive review on the subject.

Briefly, the dimer is a bipartite graph, consisting of black and white nodes, drawn on a torus, $\mathbb{T}^2$.
The edges of the dimer correspond to the fields of the SCFT and, by the bipartite condition, connect nodes of opposite color.
The nodes themselves encode the superpotential $W$:
black nodes correspond to monomial traces in $W$ with a $-1$ coefficient, while white nodes correspond to monomials of coefficient $+1$.
The monomials are then reconstructed by arranging the fields as an ordered string choosing, say, counterclockwise orientation for black nodes and clockwise for white nodes.
Finally, edges enclose faces, which in turn correspond to gauge groups of equal rank:
an edge separating face $i$ from face $j$ corresponds to a bifundamental field of the gauge groups $i$ and $j$, with the assignment of fundamental gauge group $i$ and antifundamental gauge group $j$ determined by the orientation of the embedding torus.
Such a graph can be thought of as dual to a periodic version of the familiar quiver diagram representation of the gauge theory.
The dimer model can be interpreted as a tiling~\cite{Hanany:2005ss,Franco:2005sm}, by NS-branes and D-branes, serving as a generalization of brane box models~\cite{Hanany:1997tb, Hanany:1998it}.
We illustrate the dimer with the canonical example of $\cN=4$ super-Yang--Mills theory, which corresponds to the trivial toric non-compact Calabi--Yau threefold $\IC^3$ in Figure~\ref{f:c3dimer}.

As mentioned above, the field theories under investigation flow to a non-trivial IR fixed point.
At the fixed point there is a particular $U(1)_R$ that is part of the supermultiplet that contains the stress-energy tensor.
Thus the charges under this $U(1)$, \textit{viz.}, the \textbf{$R$-charges}, of chiral operators $X$ are identified with their scaling dimensions as $\Delta[X]=\frac{3}{2}\,R[X]$.
In principle, this superconformal $R$-charge is an unknown non-anomalous combination of all the global $U(1)$ symmetries of the theory.
\textit{A priori}, accidental $U(1)$ symmetries may appear along the flow to the IR and mix with the superconformal $R$-symmetry, so one might worry that the latter is not visible in the UV theory.
However, it is known this is not the case, as potential accidental Abelian symmetries, which have been shown to appear only for theories dual to geometries with four-cycles~\cite{Benishti:2011ab}, are spontaneously broken along the flow and hence do not mix with the fixed point $R$-symmetry.
Thus, the IR $R$-symmetry for the theories of interest is in fact a combination of the non-anomalous Abelian symmetries visible in the UV Lagrangian.
As a further check, the non-anomalous orthogonal Abelian symmetries become baryonic symmetries in the IR fixed point, and it is known (\textit{e.g.},~\cite{Franco:2005rj, Martelli:2008cm}) that the number of such symmetries matches that expected from the gravity dual, hence supporting the assumption that the IR $R$-symmetry is already visible in the UV Lagrangian.

As the IR superconformal $R$-symmetry is visible in the UV, we can thus identify it using the standard $a$-maximization algorithm\footnote{
In fact, that no accidental $U(1)$s mixing with the $U(1)_R$ appear along the flow is implicitly assumed everywhere in the literature, as it is required to use the machinery of~\cite{Intriligator:2003jj}.}
and trace it back to UV.
This provides a natural choice of $R$-symmetry, and indeed we will loosely refer to the $R$-charge of a field as the $R$-charge under this particularly interesting $R$-symmetry.
It is now expedient to briefly review $a$-maximization specialized to the case of the dimer models.

\subsection{$a$-maximization for dimer models}
\label{amax}

In this subsection, let us review the $a$-maximization procedure of~\cite{Intriligator:2003jj} when implemented for our dimer models.
For conformality we need to impose the vanishing of $\beta$-functions, both for gauge coupling and superpotential couplings.
First of all, we recall that the NSVZ exact $\beta$-function for the $A$-th gauge group factor with coupling $g_A$, in terms of the $R$-charges $R_i$ of all fields $X_i$ charged under $A$, is
\begin{equation}
\beta_A = \frac{3N}{2(1-\frac{g_A^2N}{8\pi^2})}
\left[2 - \sum_{i: i \in \partial F} (1 - R_i) \right] ~.
\end{equation}
The sum runs over the sides which bound a face $F$, indicated by $i \in \pa F$, since faces correspond to gauge groups in the dimer and the edges bounding correspond to fields transforming under that group.
Note that the same field can provide two edges for a single face;
this happens when the field transforms under the adjoint representation of the corresponding gauge group.
Therefore, the imposition of the vanishing of $\beta_A$ for each $A$ requires that in the dimer, for each face,
\begin{equation}\label{conf1}
\sum_{ i : i \in \pa F } ( 1- R_i ) = 2 ~.
\end{equation}

The condition that the superpotential has $R$-charge two (or analogously the vanishing of the superpotential coupling $\beta$-function) gives the condition that, for each node $V$ in the dimer ---
which corresponds to a monomial term in the superpotential,
\begin{equation}\label{conf2}
\sum_{ i : V \in \pa ( i ) } R_i = 2 ~.
\end{equation}
The sum is over edges incident on the vertex $V$, denoted as $V\in \pa (i)$.

Then, following~\cite{Intriligator:2003jj}, subject to these constraints of the conformal manifold, one maximizes the trial $a$-function $a := \frac{3}{32}(3 \tr\,R^3 - \tr\,R)$ for a set of trial $R$-charges, where the trace indicates a 
sum over $R$ charges of the fermions (which are one less than that of 
the bosons in the same multiplet).  
For our theories, $\tr\,R = 0$, so we need only maximize
\begin{equation}\label{a}
a ( \{ R_i\} ) = \sum_{ i =1 }^d ( R_i -1 )^3 ~,
\end{equation}
where the sum is over all the $d$ edges.
There are arguments~\cite{Kato:2006vx} which say that, quite generally, the extremum is unique.

\begin{figure}[h]
\begin{center}
 \resizebox{!}{6cm}{\includegraphics{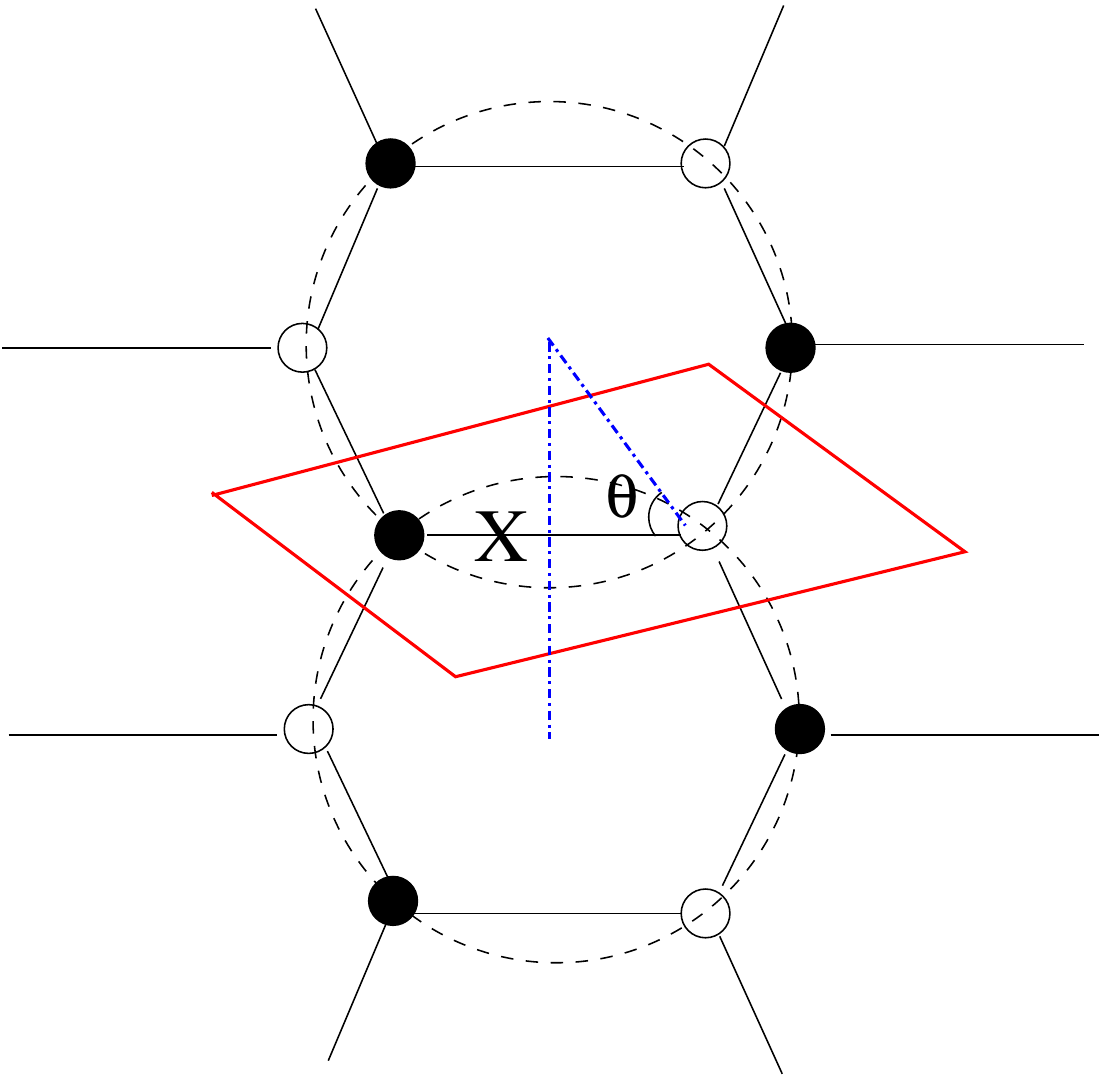}}
\caption{{\sf \small
The dimer for the $\cN=4$ super-Yang--Mills theory corresponding to the trivial CY$_3$ toric cone $\IC^3$.
The fundamental region, containing a single pair of white/black nodes --- signifying that there is only a single gauge group factor --- is marked by the parallelogram in red.
The nodes are trivalent, corresponding to two cubic monomials in the superpotential:
$W = \tr\,(XYZ - XZY)$.
We have marked one edge as the field $X$, and the angle $\theta$, in a isoradial embedding --- marked by the dotted unit circles --- is related to its $R$-charge as $\theta = \frac{\pi}{2} R(X)$.
}}
\label{f:c3dimer}
\end{center}
\end{figure}

Remarkably, the $R$-charge can be very easily encoded in the dimer.
As described in~\cite{Hanany:2005ss}, it is particularly convenient to draw the dimer in an \textbf{isoradial} embedding, a concept in fact noted in the mathematical literature (\textit{e.g.},~\cite{Kenyon-review}).
The isoradial dimer is such that all nodes lie on the circumference of a circle of unit radius centered upon each face;
whence the name isoradial.
In fact, there is a moduli space of isoradial embeddings that satisfy this condition.
Among these, a particular one, the \textbf{$R$-dimer}, is selected by the $R$-charges of the fields in the corresponding ${\cal N}=1$ quantum field theory.
As shown in~\cite{Hanany:2005ss}, the $R$-charges of a given bifundamental field $X_{ij}$, on the interface between faces $i$ and $j$, can be encoded in the angle $\theta$ subtended between the edge itself and the radius of the circle centered on face $i$ extending to the node where the $X_{ij}$ edge starts.
Indeed, because of the isoradial condition, this is the same had we chosen the circle centering face $j$.
Thus, the dimer model here is a collection of rhombi.
In summary, we have an immediate geometrical formula to read off the $R$-charge:
\begin{equation}
\theta=\frac{\pi}{2}\,R[X_{ij}] ~.
\end{equation}
We demonstrate the relevant quantities for our standard example of $\IC^3$ in Figure~\ref{f:c3dimer}.

Thus, since for a planar isoradial embedding, an $n$-sided polygon has internal angles that sum to $(n-2)\,\pi$ while all angles around a node add up to $2\pi$, the conformality conditions~\eqref{conf1} and~\eqref{conf2} are automatically guaranteed, and it remains only to maximize~\eqref{a} in terms of the angles in the dimer.

As we have reviewed, the dimer encodes all the relevant information about the field theory:
the matter content, the gauge sector, and the superpotential can all be read off from the dimer.
At this level, the precise $\mathbb{T}^2$ on which the dimer is drawn is largely irrelevant.
Crucially, the isoradial prescription, which allows us to encode the details of the IR superconformal fixed point as $R$-charges corresponding to angles, chooses a particular $\mathbb{T}^2$ for each gauge theory;
namely the $\mathbb{T}^2$ on which the isoradial dimer fits.

Since this $\mathbb{T}^2$ is fixed, it is natural to consider its properties, in particular its complex structure which we will denote as $\tau_R$.
In fact, following the isoradial prescription, and making use of the standard $a$-maximization tools, it is possible to draw the unit cell of the isoradial dimer from which $\tau_R$ is easily read off.
This process can be automated with the help of a computer.
Of course, since for any torus, the complex structure is defined only up to modular transformation, we must bear in mind that any $SL(2,\IZ)$ action on $\tau_R$ gives the same underlying $\mathbb{T}^2$ of the dimer.
This quantity $\tau_R$ will turn out to be a distinguishing property of our gauge theories.

\subsection{Seiberg duality and the invariance of $\tau_R$}

\begin{figure}[h]
\begin{center}
 \resizebox{!}{6cm}{\includegraphics{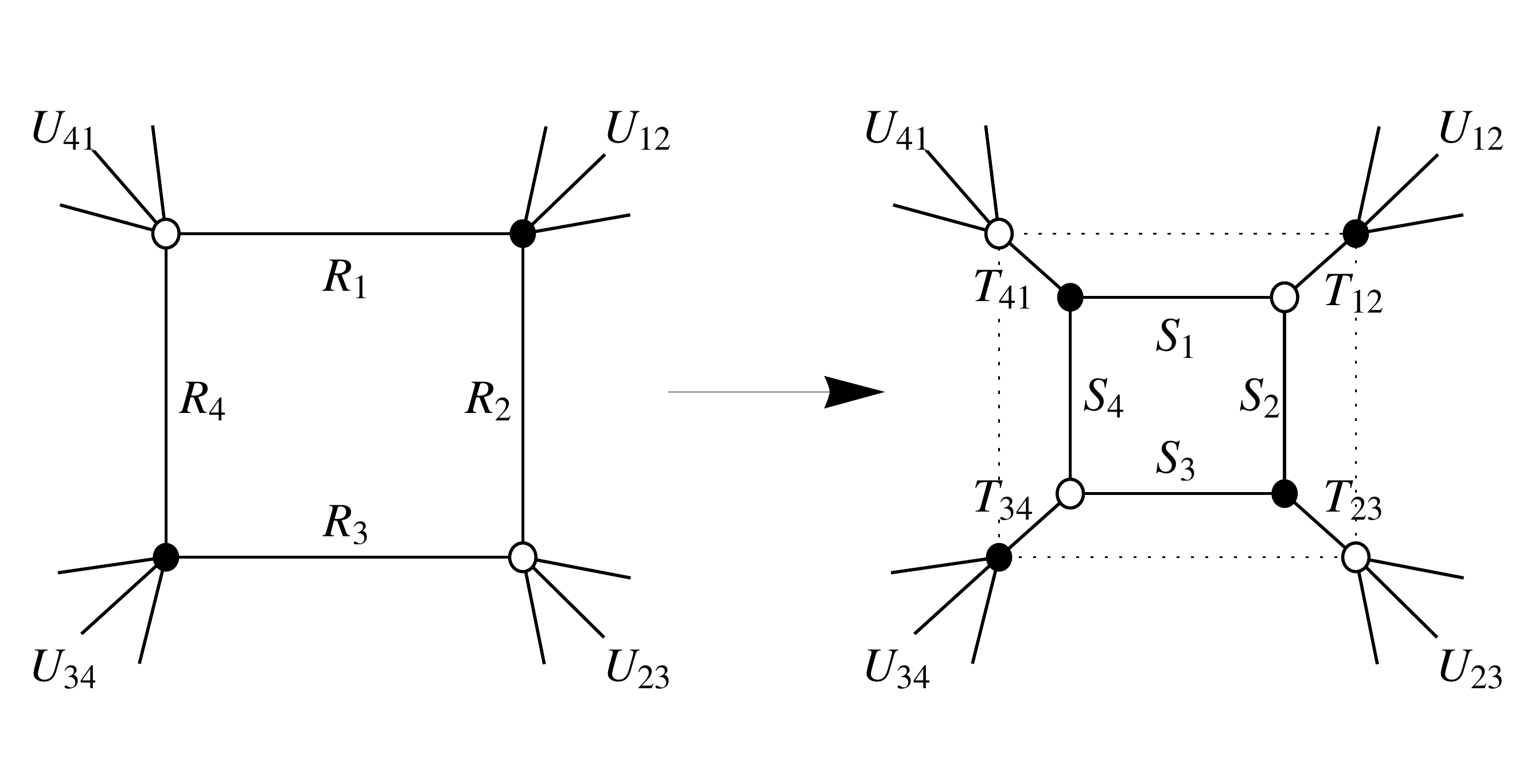}}
\caption{{\sf {\small Seiberg duality in dimer models implemented by a so called \textit{urban renewal}, together with appropriate $R$-charge assignments.
With the assumption of \textit{rigid local refinement} the four nodes on the corners of the dotted rectangle ($R$-box) remain fixed and only the position of the four new nodes ($S$-box) are to be determined.}}}
\label{fig:UR}
\end{center}
\end{figure}

We now arrive at the chief topic of our interest, Seiberg duality.
Its implementation on the field theory is well known, and we refer to~\cite{Intriligator:1995au} as the classic review.
In turn, as described in~\cite{Franco:2005rj}, Seiberg duality can be implemented in the dimer model in a very neat way, in terms of a move in the graph known as \textit{urban renewal}~\cite{Propp:urbanrenewal,Franco:2005rj}.
Rather than providing a cumbersome description in words, we provide the graphical effect of urban renewal in Figure~\ref{fig:UR}.
We note that this move, at least as a graph move, is \textit{local} in that it only alters the edges (\textit{i.e.}, fields) surrounding the face (\textit{i.e.}, gauge group) undergoing the Seiberg duality.
The face undergoing urban renewal is four sided, which in the field theory corresponds to a gauge group with $N_f=2\,N_c$ so that both electric and magnetic theories contain $SU(N_c)$ gauge groups.\footnote{
Of course, it is possible to dualize gauge groups which are not four sided.
This will not change the moduli space, but cannot be recast as a brane tiling.}

That two Seiberg (toric) dual pairs of gauge theories have their corresponding dimer models related by this urban renewal move is well known~\cite{Franco:2005rj}.
In conjunction with our distinguished isoradial embedding, we now come to the principal observation of this paper, namely that different Seiberg dual theories are encoded on dimers which live on the \textit{same} isoradial dimer.
In other words, \textit{different Seiberg dual phases of a given theory have the same $\tau_R$ up to an $SL(2,\,\mathbb{Z})$ transformation}.
More succinctly, since the Klein $j$-function is a modular invariant, we can phrase our observation as:
\begin{quote}
\fbox{Different Seiberg dual phases of a gauge theory have the same $j(\tau_R)$.}
\end{quote}

We will prove this statement in the following section.
For now, let us give ample support, which we have obtained from experimentation.
We turn to the database of dimer models compiled in~\cite{Davey:2009bp}.
Therein, we have many collections of Seiberg dual phases, including perhaps the most famous pair of the theory corresponding to the zeroth Hirzebruch surface $\IF_0 \simeq \IP^1 \times \IP^1$.
We draw the isoradial embedding of all the associated dimer models and compute their $\tau_R$;
the results are tabulated in Table~\ref{t:tau}.
The leftmost column denotes the CY$_3$ geometry which the D$3$-brane probes ($dP$ denotes cones over the del Pezzo surfaces and $PdP$, the pseudo del Pezzo ones), for bookkeeping, we record the actual $\tau_R$ values of the isoradial embedding of the various phases in the middle column, and the final column is reserved for the $j$-invariant.\footnote{
We have divided by the famous $1728$ prefactor, so we actually tabulate Klein's absolute invariant $J$.}
In Table~\ref{t:tau}, we see that while in some cases, the $\tau_R$ appear to be remarkably different, they share the same value of the $j$-function, and are thus modular equivalent.

\begin{table}
\[
\begin{array}{|c|c|c|}\hline
{\rm CY}_3 \mbox{ geometry } & \tau_R & 1728^{-1} j(\tau_R) \\ \hline \hline
\IF_0 & i, 1+i & 1 \\ \hline
L^{2,2,2} & \frac15 + \frac25 i \ , \frac15 + \frac25 i & \simeq 166.375 \\ \hline
L^{2,3,2} &
1+\frac{1+\sqrt[3]{-1} e^{\frac{2}{3} i \sqrt{7} \pi }}{1+5 e^{\frac{1}{3} i
 \left(2+\sqrt{7}\right) \pi }} \ ,
\frac{2 e^{\frac{1}{2} i \sqrt{7} \pi } \left(1+e^{-\frac{1}{3} i
 \left(\sqrt{7}-7\right) \pi }\right) \cos \left(\frac{1}{6} \left(2+\sqrt{7}\right)
 \pi \right)}{1+5 (-1)^{2/3} e^{\frac{1}{3} i \sqrt{7} \pi }}
&
\simeq -7489.12 \\ \hline
L^{2,4,2} & \frac{-3+e^{-\frac{i \pi }{\sqrt{3}}}}{\cos \left(\frac{\pi }{\sqrt{3}}\right)-3} \ ,
i \left(i+2 \tan \left(\frac{\pi }{2 \sqrt{3}}\right)+\cot \left(\frac{\pi }{2
 \sqrt{3}}\right)\right)
& 746401
\\ \hline
L^{3,3,3} & \frac{i}{3} \ , 1 + \frac{i}{3} & \simeq 88862.08 \\ \hline
dP_2 &
\begin{array}{l}
-\frac{(-1)^{9/16} e^{-\frac{1}{16} i \sqrt{33} \pi }+\sqrt[4]{-1} e^{\frac{1}{4} i
 \sqrt{33} \pi }-e^{\frac{1}{2} i \sqrt{33} \pi }+(-1)^{15/16} e^{\frac{9}{16} i
 \sqrt{33} \pi }}{i-2 \sqrt[4]{-1} e^{\frac{1}{4} i \sqrt{33} \pi }+e^{\frac{1}{2} i
 \sqrt{33} \pi }} \ , \\
\frac{\sqrt[16]{-1} \left(\sqrt[4]{-1}-e^{\frac{1}{4} i \sqrt{33} \pi
 }\right)^2}{(-1)^{9/16}+(-1)^{5/8} e^{-\frac{1}{16} i \sqrt{33} \pi }-(-1)^{5/16}
 e^{\frac{1}{4} i \sqrt{33} \pi }-e^{\frac{9}{16} i \sqrt{33} \pi }}
\end{array}
& \simeq 0.169478 \\ \hline
dP_3 & e^{\pi i/3} \ , e^{\pi i/3} \ , e^{\pi i/3} \ , e^{\pi i/3} & 0 \\ \hline
PdP_3b & 1+i \cos \left(\frac{\sqrt{5} \pi }{2}\right) \csc \left(\frac{3 \sqrt{5} \pi
 }{2}\right) \ ,
1+i \cos \left(\frac{\sqrt{5} \pi }{2}\right) \csc \left(\frac{3 \sqrt{5} \pi
 }{2}\right)
& \simeq 1.02155 \\
\hline
\end{array}
\]
\caption{{\sf \small
The values of $\tau_R$ for different Seiberg dual faces of a theory are $SL(2,\mathbb{Z})$ equivalent.
}}
\label{t:tau}
\end{table}

The cases where $j(\tau_R) = 0$ and $j(\tau_R) = 1728$ correspond to elliptic curves with enhanced symmetries, $\mathbb{Z}_2\times \mathbb{Z}_3$ and $\mathbb{Z}_4$, respectively, whereas for a generic value of $j(\tau_R)$, the elliptic curve $y^2 = x(x-1)(x-\lambda)$ only enjoys a $\mathbb{Z}_2$ symmetry corresponding to the invariance under $y\mapsto -y$.

\section{A proof of $j(\tau_R)$ as a Seiberg duality invariant}\setall
\label{sec:UrbanRen}

We now present a proof of the claim in the previous section.
To that end, we need to develop a finer understanding of the Seiberg duality procedure implemented on the dimer as an urban renewal move.
As usual, we will adhere to the terminology that the theory before Seiberg duality/urban renewal, is \textit{electric} and the one after is \textit{magnetic}.

\subsection{A closer look at urban renewal}

As a graph theoretic operation, Seiberg duality is a local change in the connectivity of the dimer described by urban renewal as depicted in Figure~\ref{fig:UR}.
However, as discussed above, the isoradial prescription endows the angles and lengths of the isoradial dimer with a special significance.
Thus, in principle, performing the urban renewal move on an $R_e$-dimer --- the electric theory --- yields another graph on which we should run independently the $a$-maximization procedure, \textit{a priori} yielding another $R_m$-dimer for the magnetic theory.

While the $R_m$-dimer will have the connectivity dictated by the urban renewal move, it might, in principle, have all angles and lengths different from the initial $R_e$-dimer.
Let us assume momentarily however that this is not the case, and that in fact urban renewal is indeed a local change.
More precisely, we consider the urban renewal move shown in Figure~\ref{fig:UR} and label the edges of the original square $1,2,3,4$ and the corresponding $R$ charges $R_1, R_2, R_3, R_4$ ---
we will call this the $R$-box.
The shrunken square has $R$-charges $S_1, S_2, S_3, S_4$.
Let the $R$-charges of the diagonals be $T_{41}, T_{12}, T_{23}, T_{34}$ as shown in the figure.
In the corresponding field theory, the edges with $R$-charges $R$, $S$, and $T$ correspond to quarks, dual quarks, and Seiberg mesons, respectively.

We now assume that a \textbf{rigid local refinement} exists, that is, that there is a consistent set of $R$-charge assignments compatible with $a$-maximization so that the $S$-$T$ refinement of the $R$-box can be drawn by urban renewal \textit{without changing the vertices of the $R$-box} while leaving all other $R$-charges outside the box untouched.
In other words, fields untouched by Seiberg duality retain their $R$-charge.
We will discuss this assumption in detail later and for now, our immediate goal is to show how such refined $R$-charges are found without incurring any contradiction.

We proceed then with the assumption that the charges on both left- and right-hand side of Figure~\ref{fig:UR} provide a valid isoradial embedding and find the relations that must be satisfied.
Basically, there will be a linear constraint for every node and face involved, and that will be enough to solve for the eight unknowns $S_i$ and $T_{ij}$ in terms of the original $R_i$.

We first impose the consistency conditions, using~\eqref{conf2}, from the trivalent vertices that total $R$-charge at each node is two.
These give
\begin{equation}
\label{trivalentconds}
\begin{split}
& S_1 + S_2 + T_{12} = 2 ~; \\
& S_2 + S_3 + T_{23} = 2 ~; \\
& S_3 + S_4 + T_{34} = 2 ~; \\
& S_4 + S_1 + T_{41} = 2 ~.
\end{split}
\end{equation}
In the corresponding field theory, these four nodes correspond to the four cubic potentials that arise in Seiberg duality as the interaction between Seiberg mesons and dual quarks.
Similar constraints arise from the consistency conditions from the corners of the $R$-box
\begin{equation}
\label{cornersafter}
\begin{split}
U_{41} + T_{41} &= 2 ~; \\
U_{12} + T_{12} &= 2 ~; \\
U_{23} + T_{23} &= 2 ~; \\
U_{34} + T_{34} &= 2 ~,
\end{split}
\end{equation}
where $U_{ij}$ is the total $R$-charge for extra fields at the $(ij)$ corner.

Before the refinement by urban renewal the conditions were
\begin{equation}
\label{cornersbefore}
\begin{split}
U_{12} + R_1 + R_2 & = 2 ~; \\
U_{23} + R_2 + R_3 & = 2 ~; \\
U_{34} + R_3 + R_4 & = 2 ~; \\
U_{41} + R_{4} + R_1& = 2 ~.
\end{split}
\end{equation}

Combining the corner conditions (\ref{cornersbefore}) and (\ref{cornersafter}), we deduce
\begin{equation}
\label{TRconds}
\begin{split}
T_{12} &= R_1 + R_2 ~; \\
T_{23} &= R_2 + R_3 ~; \\
T_{34} &= R_3 + R_4 ~; \\
T_{41 }&= R_1 + R_4 ~.
\end{split}
\end{equation}
This follows from the property that Seiberg mesons are constructed as bilinears of quarks, and being chiral operators, their $R$-charges add up.
This fully determines the $T_{ij}$ charges in terms of the starting charges $R_i$.
It remains to find the $S_i$'s.

We have so far made use of the conditions on the vanishing of the $\beta$-function for superpotential couplings.
We now turn to the conditions arising from setting the Yang--Mills coupling $\beta$-functions to zero, coming to \eqref{conf1}.
To that end, we recall that the angle subtended by an edge $i$ from the center of a face is $\pi ( 1 - R_i )$, while the $\beta_{A}=0$ conditions is summarized in that these angles at the center sum up to $2 \pi$.
Thus,
\begin{equation}
2\pi = \sum_i \pi (1 - S_i ) = 4\pi - \pi \sum_i S_i ~.
\end{equation}
The condition from the $S$-box is then
\begin{equation}
\label{Sface}
S_1 + S_2 + S_3 + S_4 = 2 ~.
\end{equation}
Similarly, the condition from the $R$-box is
\begin{equation}
\label{Rface}
R_1 + R_2 + R_3 + R_4 = 2 ~.
\end{equation}

Next, observe that the angle subtended by the $R_1$ from the center of the $i$-th outer face is equal to the sum of the angles subtended by $T_{41}, S_1, T_{12}$.
So we have
\begin{equation}
( 1 - R_1 ) = ( 1- T_{41} ) + ( 1 - S_1 ) + ( 1 - T_{12} ) ~.
\end{equation}
This relation can be seen in Figure~\ref{fig:UR} as arising from the $\beta$-function for the face \textit{above} the $R$-box on the left-hand side and the $S$-box on the right-hand side.
By treating $R_2, R_3, R_4$ in a similar fashion, we find
\begin{equation}
\label{rigidity}
\begin{split}
( 1 - R_2 ) &= ( 1- T_{12} ) + ( 1 - S_2 ) + ( 1 - T_{23} ) ~, \\
( 1 - R_3 ) &= ( 1- T_{23} ) + ( 1 - S_3 ) + ( 1 - T_{34} ) ~, \\
( 1 - R_4 ) &= ( 1- T_{34} ) + ( 1 - S_4 ) + ( 1 - T_{41} ) ~.
\end{split}
\end{equation}

Combining these with equations (\ref{TRconds}), we obtain
\begin{equation}
\label{SR}
\begin{split}
S_1 &= 2 - R_1 - R_2 - R_4 = R_3 ~, \\
S_2 &= 2 - R_1 - R_2 - R_3 = R_4 ~, \\
S_3 &= 2 - R_3 - R_2 - R_4 = R_1 ~, \\
S_4 &= 2 - R_4 - R_3 - R_1 = R_2 ~,
\end{split}
\end{equation}
which is the final result for $S_i$ charges.

Physically, the meaning of these relations comes from the brane realization of these models.
Seiberg duality in brane intervals is realized as the exchange of two NS-branes.
Here we have a two-dimensional generalization of the same phenomenon.
The urban renewal of Figure~\ref{fig:UR} can be interpreted as the simultaneous exchange of the upper and lower edges together with the left and right edges ---
in other words, the simultaneous exchange of opposite edges.
Thus the edge with $R$-charge $R_2$ becomes the edge with $R$-charge $S_4$, etc.
The computation above shows that the edges in fact keep their $R$-charges along this transition.

One can easily check that these solutions (\ref{TRconds}), (\ref{SR}) for the $S_i$ and $T_{ij}$ ---
$R$-charges after the urban renewal ---
solve the full set of linear constraints after the renewal (\ref{trivalentconds}), (\ref{Sface}), (\ref{rigidity}), given that charges $R_i$ before the procedure also satisfied constraints.

So far we have implemented the constraints arising from the vanishing of the $\beta$-functions of both superpotential and Yang--Mills couplings, resulting in a set of new $R$-charges given by (\ref{SR}) for the new fields.
It remains to be proven that this new assignment is consistent with the maximization of the central charge $a$.
As the untouched parts of the dimer remain the same as before the move, we can concentrate on the correction to the central charge $a$ due to the urban renewal move.
Therefore, the change in the $a$-function is
\bea
\Delta a = \sum_{i=1}^4 (S_i-1)^3 + (T_{12}-1)^3 + (T_{23}-1)^3 + (T_{34}-1)^3 + (T_{41}-1)^3 - \sum_{i=1}^4 (R_i-1)^3 ~,
\nn \\
\eea
and upon using the relations \eqref{TRconds}, \eqref{SR} and using \eqref{Rface}, we have that
\begin{equation}
\Delta a=0 ~.
\end{equation}
The central charge of the theory arising upon urban renewal, with the $R$-charge assignment given by (\ref{SR}), coincides with that of the original theory, as necessary for a Seiberg dual pair.

A comment is in order here.
The $R$-charges of the edges are encoded in terms of $R$-charges $a_i$ that are assigned to external points in the toric diagram (we will say more about the $a_i$ charges and the following construction in Section~\ref{sec:taufromtoric}, here just mentioning the main facts).
Each $R$-charge is a linear combination of the $a_i$ with coefficients $1$ or $0$.
The way this is computed is by looking at the two zig-zag paths that pass through the edge.
There are precisely two such paths.
Each path maps to a $(p,q)$-leg in a $(p,q)$-web that is dual to the toric diagram.
Using this, we find that every edge gives rise to a wedge that is spanned by two $(p,q)$-legs.
The linear combination of $a_i$ is such that all points in the wedge contribute with a coefficient $1$ and all points outside the wedge contribute with a coefficient $0$.
What we can learn from this fact is that the central charge can be encoded in terms of the charges $a_i$ only without making any use of the specific brane tiling which is used, be it the one before or after urban renewal.
Since the $a$-maximization is performed on any set of variables, one can perform the maximization on $a_i$, which depend on the toric diagram but not on the specific brane tiling.
As a result, it is enough to show that the central charge $a$ is unchanged in order to argue that the maximum remains the same for both phases.

In summary, we have started with a dimer endowed with an isoradial embedding prescribed by $R$-charges that satisfy $a$-maximization.
Using it we have made a trial assignment of $R$-charges for the Seiberg dual dimer, given by (\ref{TRconds}), (\ref{SR}) for the modified box and all other charges remaining unchanged.
Then we saw that the new charges not only satisfy all linear constraints, but also give the same value for $a$ as the original dimer.
Since the new dimer, being Seiberg dual, should have the same \textit{maximum} value of $a$ as the original one, this, in fact, shows that our trial assignment of $R$-charges maximizes $a$.
This almost proves the statement of $j(\tau_R)$ invariance, because our trial assignment did leave the global structure of the tiling unchanged.
Before we conclude that, however, we need to clarify a couple of subtleties.

\subsubsection{Integrating out massive fields}

\begin{figure}[ht]
\begin{center}
 \resizebox{!}{6cm}{\includegraphics{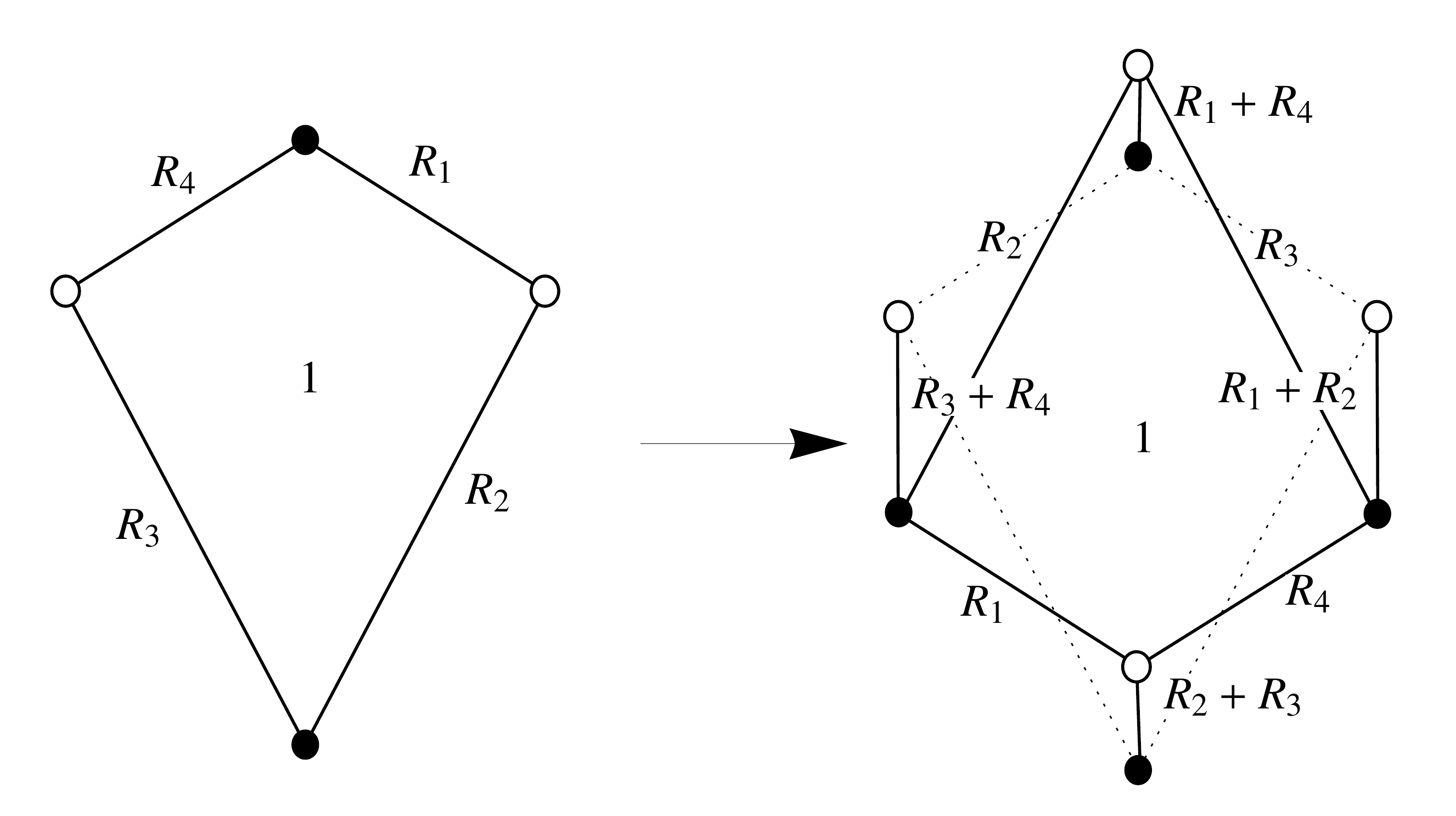}}
\caption{{\sf {\small Urban renewal taking into account linear constraints.}}}
\label{fig:URconstrained}
\end{center}
\end{figure}

Inspection of $S_i$ solutions (\ref{SR}) shows that the intuitive picture of urban renewal in Figure~\ref{fig:UR} is, in fact, a bit misleading.
Namely, we find that $S_i=R_{i+2}$, so the dimer cannot become smaller in isoradial embedding, it just gets flipped and offset.
A more realistic picture is shown in Figure~\ref{fig:URconstrained}.
This means that the dimer inevitably will go on top of some of the original nodes (unless all $T_{ij}=1$, so the corresponding lengths are zero and the dimer remains in the same place).
This seems worrisome ---
how will we ever get a sensible dimer (tiling) then?

\begin{figure}[ht]
\begin{center}
\resizebox{!}{4cm}{\includegraphics{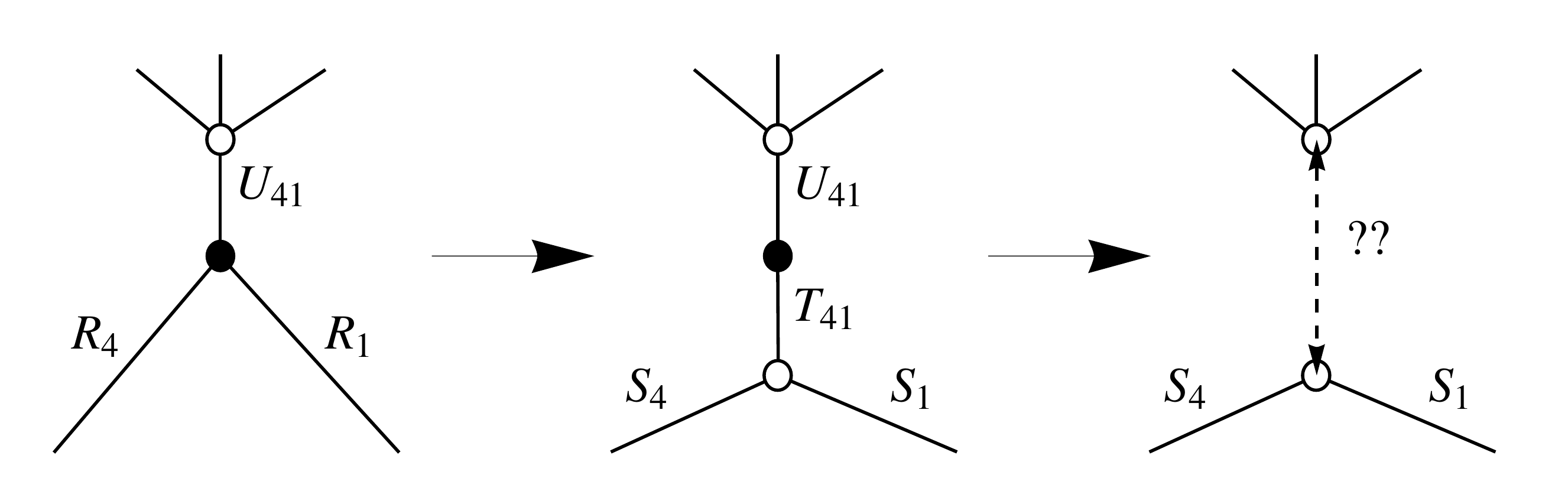}}
\caption{{\sf {\small Urban renewal procedure focusing on a three-valent corner of the $R$-box.
Insertion of $T_{41}$ causes a two-valent node to appear, which must be removed by integrating out a massive field.
This looks like a potential problem to collapse the two white nodes into one.}}}
\label{fig:urban2valent1}
\end{center}
\end{figure}

This issue is related to another step in the urban renewal procedure that we glanced over, namely the integration of massive fields.
If, after urban renewal, we end up with any two-valent nodes in the dimer, that signals the appearance of massive fields, which must be integrated out.
This is a well known procedure, which from the dimer perspective removes the two-valent node, and collapses two adjacent nodes into one.
The situation always occurs if a corner of the $R$-box before urban renewal was a three-valent node, as depicted in Figure~\ref{fig:urban2valent1}.
This might seem even more problematic, because at a first glance it is not clear how we can collapse the two nodes adjacent to the two-valent node without moving everything in the isoradial embedding.

\begin{figure}[ht]
\begin{center}
\resizebox{!}{4cm}{\includegraphics{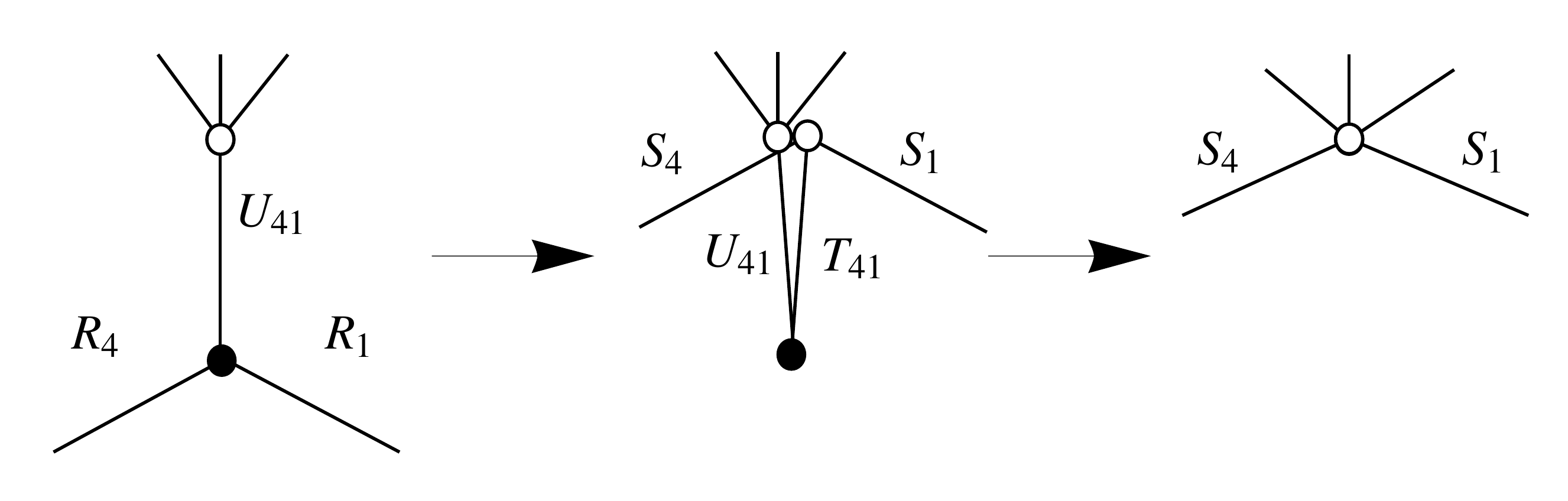}}
\caption{{\sf {\small Urban renewal and integration out of a massive field in isoradial embedding.
The two nodes that need to be collapsed are already \textit{in the same position}, as in the middle picture, and nothing needs to be moved.}}}
\label{fig:urban2valent2}
\end{center}
\end{figure}

The answer to the puzzle comes from looking more carefully at the linear constraints.
We have already checked in the previous section, that all the charges can be assigned consistently after the urban renewal procedure (but before integration of massive fields).
This means that in the situation depicted in Figure~\ref{fig:urban2valent1} the charges of fields connected to the two-valent node satisfy:
\begin{equation}
U_{41}+T_{41} = 2 ~.
\label{UTrelation}
\end{equation}
The lengths of the corresponding edges in isoradial embedding are then:
\begin{equation}
l_{U_{41}} = 2 \cos\left(\frac{\pi U_{41}}{2}\right) = 2 \cos\left(\frac{\pi (2-T_{41})}{2}\right) = - 2 \cos\left(\frac{\pi T_{41}}{2}\right) = -l_{T_{41}} ~,
\end{equation}
while the angle in between is
\begin{equation}
\theta = \pi \frac{T_{41}+U_{41}}{2} = \pi ~.
\end{equation}
Negative lengths are fine from the point of view of isoradial embedding;
this just means that the edge extends in the opposite direction.
These relationships imply that the two nodes adjacent to the two-valent node will actually be at exactly the same position in isoradial embedding!
A more accurate depiction of the situation is as shown in Figure~\ref{fig:urban2valent2}.
This means that there is no problem with removing the two-valent node and collapsing the two ---
nothing else in the isoradial embedding needs to be moved, and all $R$-charges can be kept unchanged.

One final point is that the process of getting rid of two edges incident on one edge also conserves the $a$-function, because, using~\eref{UTrelation},
\bea
(U_{41}-1)^3 + (T_{41}-1)^3 = (U_{41}-1)^3 + (2-U_{41}-1)^3 = 0 ~.
\eea
That means in the end, after urban renewal and integration out of the massive fields, we still end up with an assignment of $R$-charges that satisfies $a$-maximization.

\begin{figure}[ht]
\begin{center}
\resizebox{!}{5cm}{\includegraphics{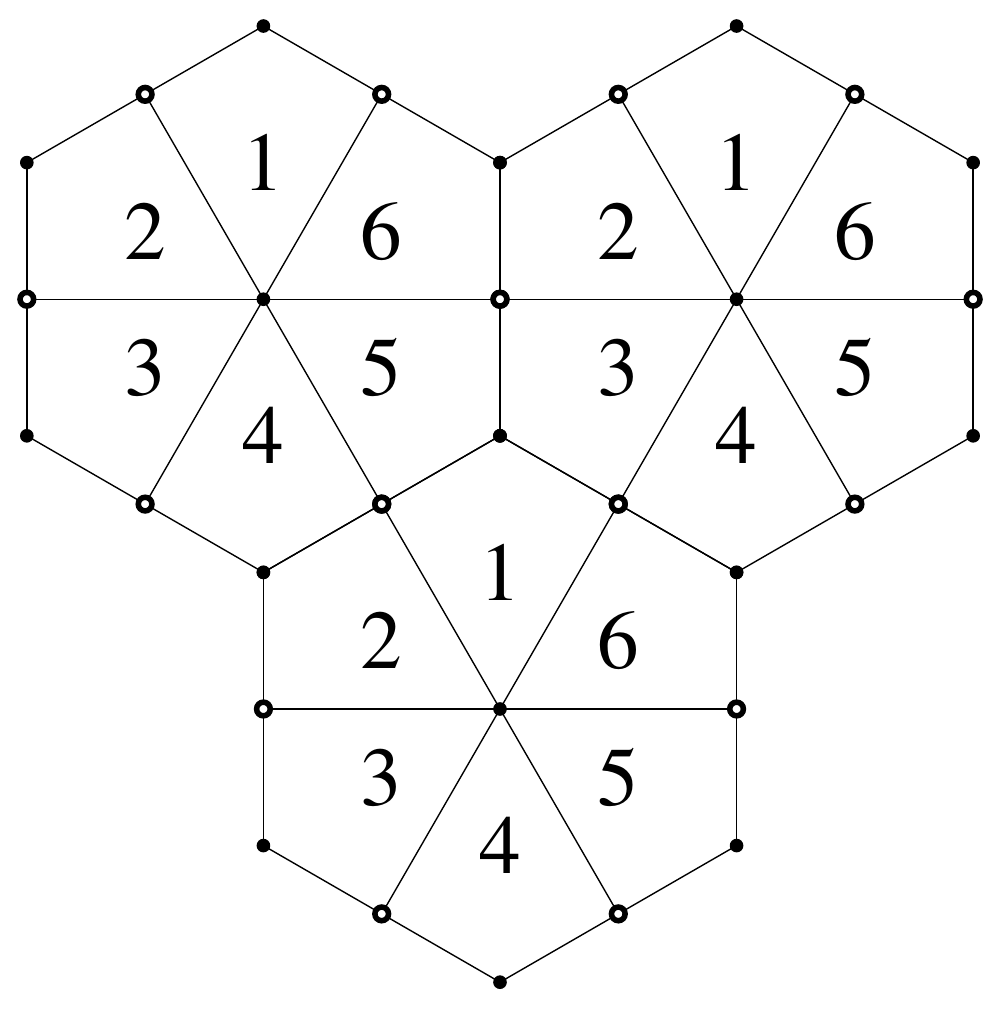}}
\resizebox{!}{5cm}{\includegraphics{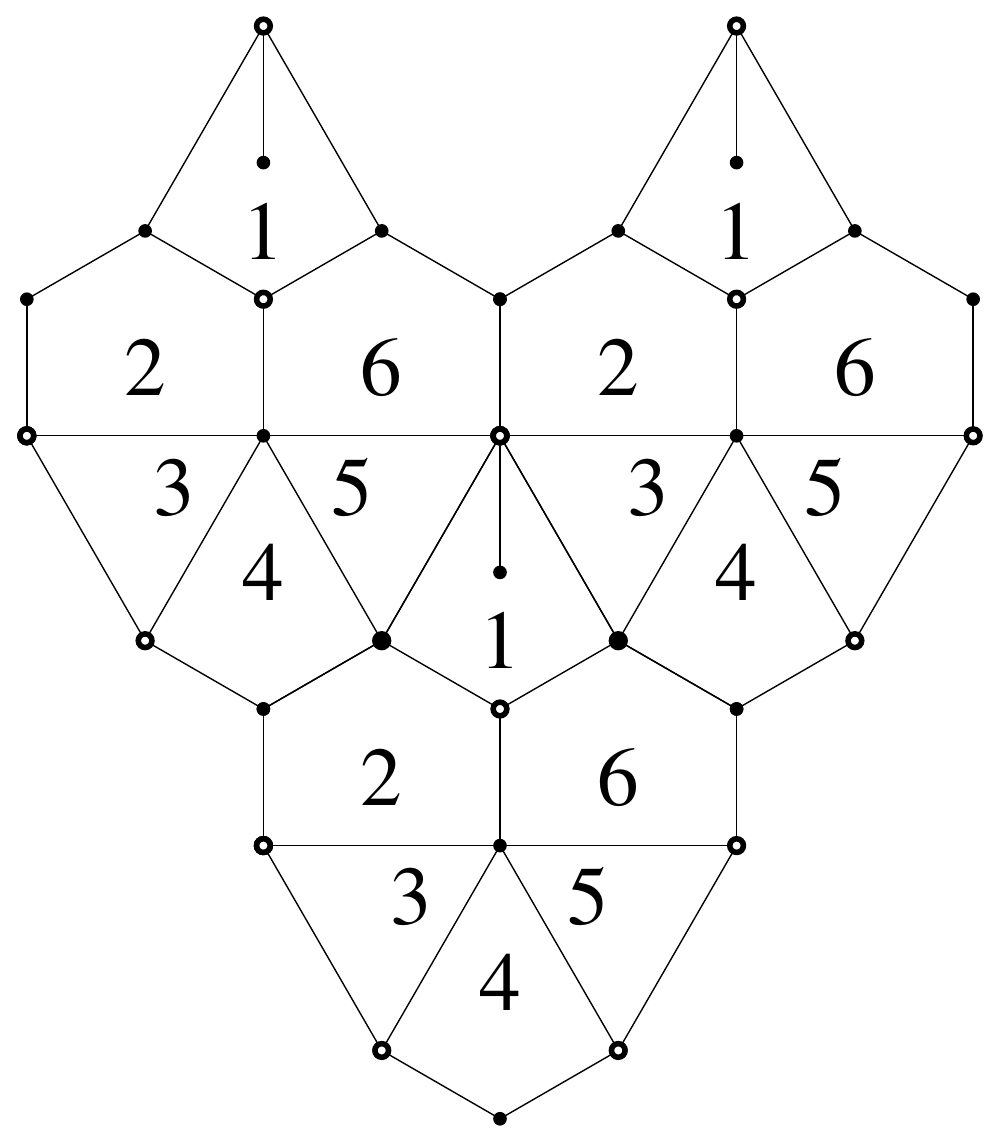}}
\resizebox{!}{5cm}{\includegraphics{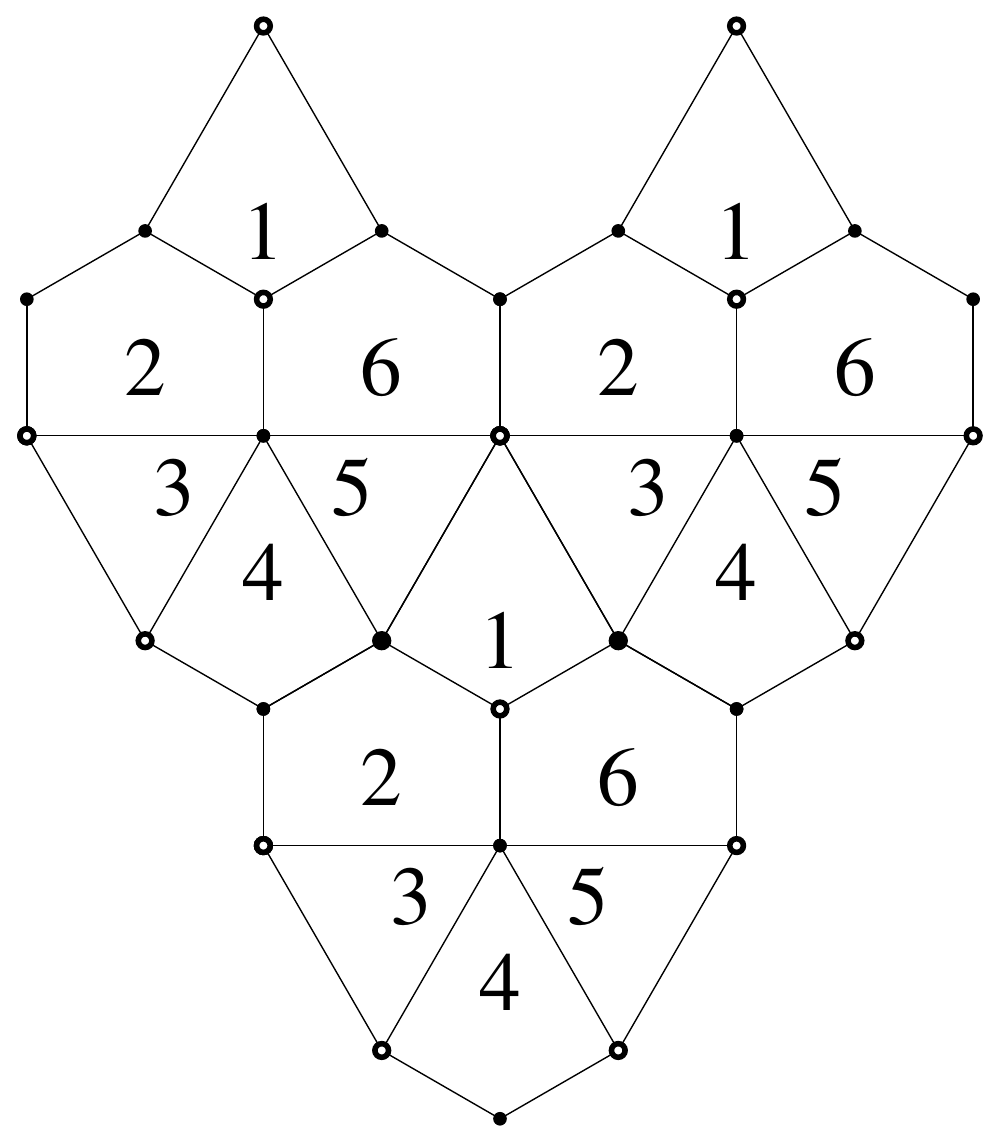}}
\caption{{\sf {\small Seiberg duality between phases I and II of $dP_3$.
Urban renewal is performed on face one of the tiling.
The first step flips and moves the tile and second step integrates out a massive field thus introduced.
Note two of the new fields have length zero ($R=1$) and are not visible in isoradial embedding.}}}
\label{fig:URdp3}
\end{center}
\end{figure}

For a concrete example, we illustrate the full urban renewal process in the isoradial embedding of the dimer corresponding to $dP_3$, the Calabi--Yau cone over the third del Pezzo surface, or $\IP^2$ blown up at three generic points.
This is drawn in Figure~\ref{fig:URdp3}.

\subsubsection{Physics of  the assumption of local rigidity }

So far we have found a consistent assignment of $R$-charges after Seiberg duality, implemented on the dimer through the urban renewal move, under the crucial assumption of local rigid refinement:
that the $R$-charges of the fields untouched by Seiberg duality remain unchanged.
As promised, we now return to examine this assumption in detail.

As a Seiberg dual pair, the original and the urban renewed theory will necessarily have the same mesonic moduli space and the same central charge.
The assumption of local rigidity, as shown above, 
  fulfills these expectations.
This assumption fixes the $R$-charges of some of the fields of the magnetic theory ---
namely those untouched by Seiberg duality ---
to be the same as in the original electric theory.
This means that the magnetic trial $R$-charge is not the most general one.

In order to further understand this point, let us consider the fields in the electric theory, which are untouched by the Seiberg duality and thus are mapped trivially into the magnetic theory.
These fields will be either adjoints or bifundamentals of the gauge group factors which are also trivially mapped by Seiberg duality into the magnetic theory.
Let us denote these fields as $X^a_{ij}$, with $i,\, j$ running through those gauge groups on which Seiberg duality is not performed and $a$ an index keeping track of their multiplicity.

Clearly, we can construct the gauge invariant chiral operators in the electric theory with baryons $\mathcal{B}[X_{ij}^a]={\rm det}X_{ij}^a$ if $i\ne j$ and mesons $\mathcal{M}[X_{ij}^a]={\rm Tr}\,X_{ii}^a$ if $i=j$.\footnote{
We might imagine constructing more gauge invariant chiral operators, but for our purposes it is sufficient to just considering these.}
By construction, these operators are trivially mapped into the magnetic theory.
On the other hand, Seiberg duality demands that the chiral ring and the set of baryonic operators are isomorphic.
This means that the baryonic operator $\mathcal{B}[X_{ij}^a]$ must have, in both theories, the same dimension and be charged under the same baryonic symmetry.
As $\Delta[B[X_{ij}^a]=\frac{3\,N}{2}\,R[X_{ij}^a]$ in both theories, it is obvious that $R[X_{ij}^a]$ must be the same in both the electric and magnetic theories.
Similarly, this holds for the mesonic operators $\mathcal{M}[X_{ii}^a]$, so that $R[X_{ii}^a]$ also must be equal in the two theories.

We have argued that the $R$-charges of fields trivially mapped under Seiberg duality must remain the same at the fixed point.
We have also explained that for the theories of interest there is no mixing with accidental Abelian symmetries along the flow, and so it is consistent, for fields trivially mapped under Seiberg duality, to choose in the UV trial $R$-charges that are the same as in the original electric theory.
This explains the assumption of local rigidity.

\subsection{$j(\tau_R)$ is an invariant under Seiberg duality}

We can now come back to our original problem of the invariance of the $\tau_R$ under Seiberg duality.
We have just proved that urban renewal is indeed a local movement in the isoradial dimer.
As such, the global structure of the dimer does not change.
In particular, the unit cell of the torus where the $R$-dimer is drawn will not change, which immediately implies that the \textit{$\tau_R$ is kept invariant --- up to an $SL(2,\mathbb{Z})$ transformation --- under Seiberg duality}.

In Section~\ref{sec:taufromtoric} we shall establish the invariance of $\tau_R$ by considering only the toric diagram.
These two sections provide a purely field theoretic argument for the invariance of $\tau_R$ among Seiberg dual phases.
However, it is interesting to explore the string theoretic perspective on this.
We will postpone such an analysis to a forthcoming publication~\cite{SLAGS}.

\section{$\tau_R$ from toric data}
\label{sec:taufromtoric}

In Section~\ref{sec:UrbanRen}, we have analyzed the urban renewal procedure and shown that the global structure of $R$-dimer is unchanged, which keeps $j(\tau_R)$ invariant between Seiberg dual phases.
In this section we take one more step and show how it is possible to compute $\tau_R$ directly from the toric diagram, bypassing the construction of specific dimers altogether.
For this purpose we will use the formulation of $a$-maximization based purely on the toric data as in~\cite{Butti:2005vn, Kato:2006vx}.

Note that by the arguments of this section the invariance of $j(\tau_R)$ would appear trivial, because no information about the dimer is used in the calculation, only the toric diagram.
However, there seems to be some uncertainty~\cite{Yamazaki:2008bt}, as to whether the methods of~\cite{Butti:2005vn, Kato:2006vx} apply to any phase of the theory or just the so-called \textit{minimal phase}, which is the one with the fewest perfect matchings in the brane tiling.
In this light, our explicit comparison of phases in Section~\ref{sec:UrbanRen} gives us encouragement to propose that the calculation in this section is, in fact, valid for \textit{any} phase.
We will attempt to further clarify this point in the following discussion.

For examples of $\tau_R$ calculation using the methods of this section see Appendix~\ref{app:tauexamples}.

\subsection{$a$-maximization from toric diagram}

\begin{figure}[h]
\begin{center}
 \resizebox{!}{6cm}{\includegraphics{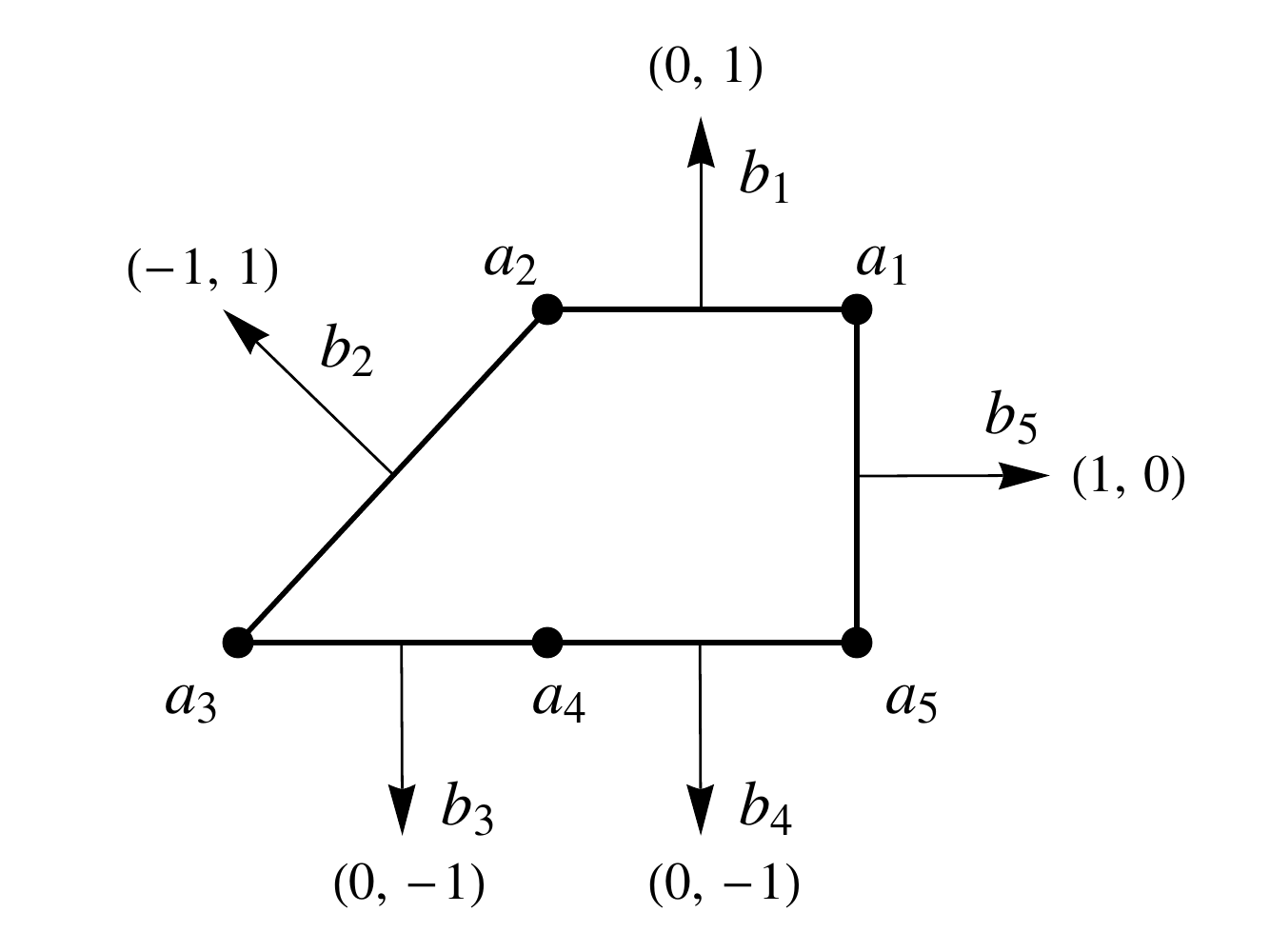}}
\caption{{\sf \small
The toric diagram for SPP.
Each external point has a variable $a_i$ assigned, and each primitive normal has variable $b_i$.
The $(p_i, q_i)$ numbers of primitive normals indicate the winding numbers of the corresponding zig-zag paths.}}
\label{fig:spptoric}
\end{center}
\end{figure}

\begin{figure}[h]
\begin{center}
 \resizebox{!}{6cm}{\includegraphics{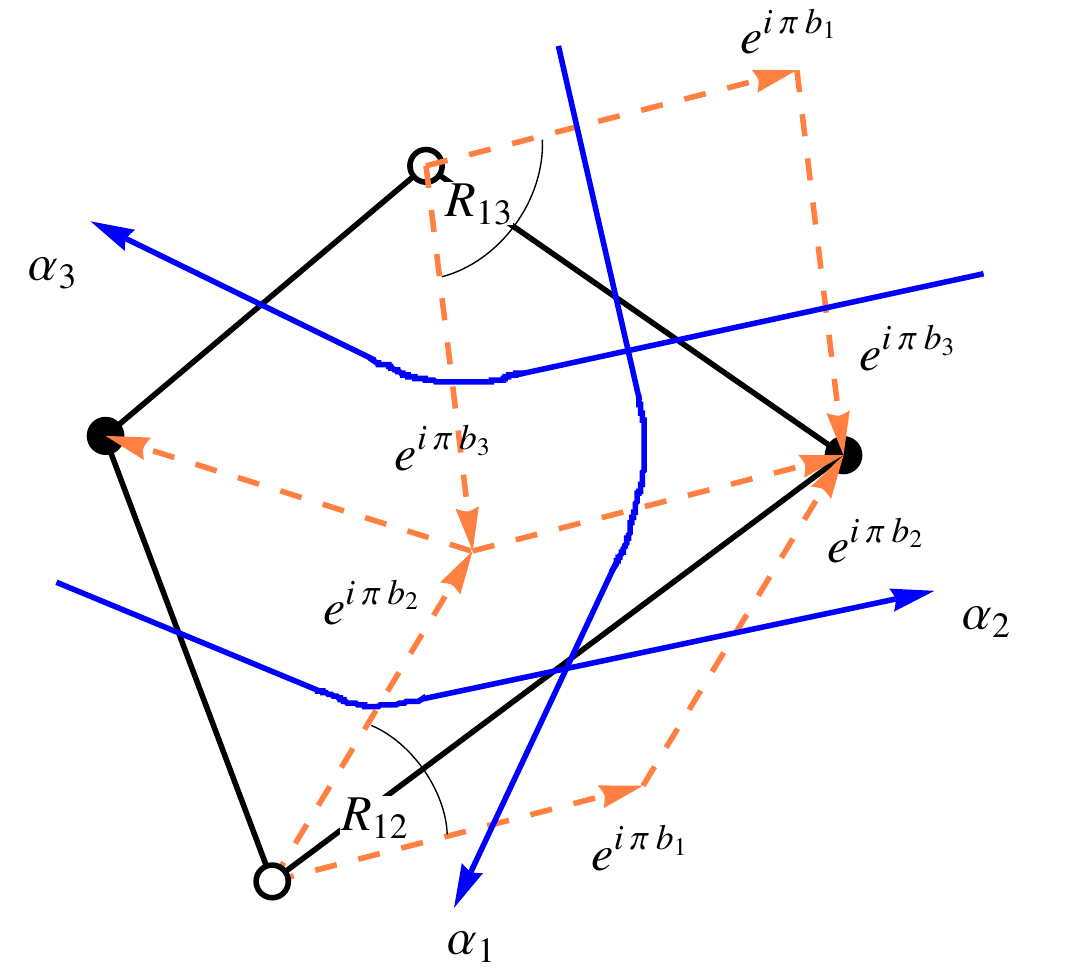}}
\caption{{\sf \small
Zig-zag paths and rhombi.
Blue lines $\alpha_i$ indicate zig-zag paths and orange lines are rhombi edges with assigned direction.
The vectors of all rhombi edges intersected by the same $\alpha_i$ are equal and parameterized as $\exp(i\pi b_i)$.}}
\label{fig:zigzags}
\end{center}
\end{figure}

First, we use the method of~\cite{Butti:2005vn, Kato:2006vx} to write the trial $a$-function directly from the toric diagram.
In order to do that we assign a variable $a_i$ to each external point of the toric diagram (see Figure~\ref{fig:spptoric}).
There is a single constraint
\begin{equation}
	\sum_{i=1}^{t} a_i = 2
\end{equation}
leaving $t-1$ independent variables, where $t$ is the number of external points in the toric diagram.
This is the number of non-anomalous $U(1)$ symmetries and thus the expected number of variables for $a$-maximization.
The geometric meaning of these variables is more clear if we consider corresponding variables for the primitive normals:
\begin{equation}
\label{eq:b_from_a}
\begin{split}
	b_1 &= a_1 ~, \\
	b_2 &= a_1 + a_2 ~, \\
	b_i &= \sum_{j=1}^{i} a_i ~, \\
	b_t &= 2 ~.
\end{split}
\end{equation}
That is, $a_i = b_{i} - b_{i-1}$ and $b_i$ are only defined up to addition of an overall constant.
If we consider a zig-zag path $\alpha_i$ corresponding to the normal, then $\pi b_i$ is the angular direction of the rhombi edges that the zig-zag path intersects (see Figure~\ref{fig:zigzags}).
The construction relies on the fact that a zig-zag path always crosses rhombi through opposite edges, which are parallel, and so \textit{all} the edges that the zig-zag path crosses are parallel with direction parameterized by $b_i$.
Note that we assign a direction to the rhombi edge such that the arrow points to the left looking down the zig-zag path direction.
Finally, it will be useful to think of the plane of the tiling as the complex plane, in which case the vectors corresponding to the rhombi edges, all having length one, are written as $e^{i \pi b_i}$.

Now, given the parameterization, the trial $a$-function is constructed as follows.
Every edge of the tiling is at an intersection of two zig-zag paths $\alpha_i$, $\alpha_j$.
If we adopt a convention that on the intersection $\alpha_j$ is counter-clockwise from $\alpha_i$, then the $R$-charge of the corresponding field, as can be easily read off from the diagram, is
\begin{equation}
\label{eq:R_from_b1}
\begin{split}
	R_{ij} &= b_j - b_i = a_{i+1} + a_{i+2} + \ldots + a_j ~.
\end{split}
\end{equation}
We have to be careful when we go around the diagram, and since in our current convention (\ref{eq:b_from_a}) $b_i$ are increasing, we have to take for $j<i$:
\begin{equation}
\label{eq:R_from_b2}
	R_{ij} = 2 + b_j - b_i = a_{i+1} + \ldots + a_t + a_1 + \ldots + a_j ~.
\end{equation}
Having parameterized all field charges in terms of $b_i$ and knowing that for every intersection of the zig-zag paths there is a corresponding field, it is easy to write the $a$-function~\cite{Butti:2005vn, Kato:2006vx}:
\begin{equation}
\label{eq:atrial}
	a(b_i) = \frac{9}{32}\left(F + \sum_{i,j,\; \langle w_i, w_j \rangle > 0 } \langle w_i, w_j \rangle \; (R_{ij} - 1)^3 \right) ~,
\end{equation} where $R_{ij}$ is (\ref{eq:R_from_b1}) or
(\ref{eq:R_from_b2}) depending on whether $i<j$.
Here $w_i=(p_i, q_i)$ are the primitive normal vectors of the toric diagram and
\begin{equation}
	\langle w_i, w_j \rangle \equiv {\rm det} \left( \begin{array}[pos]{cc} p_i & q_i \\ p_j & q_j \end{array} \right) ~.
\end{equation}
The sum is performed only over $\langle w_i, w_j \rangle > 0$ meaning $w_j$ is counter-clockwise from $w_i$.
So, maximizing this $a(b_i)$ as a function of $t-1$ variables we get the preferred isoradial embedding and the true $R$-charges for all fields using (\ref{eq:R_from_b1}), (\ref{eq:R_from_b2}).

Before continuing on, let us briefly discuss how dual phases are related from the perspective of this construction.
The difference turns out to be in how many times any two zig-zag paths $\alpha_i, \alpha_j$ actually intersect.
Given their winding numbers $w_i, w_j$, we know that they have to intersect \textit{at least} $\langle w_i, w_j \rangle$ times, but it could be more.
For example, it can be seen that during urban renewal four new intersections are introduced between zig-zag paths\footnote{
See, for example, Figure~25 of~\cite{Krippendorf:2010hj}.}
corresponding to four new fields in Figure~\ref{fig:UR}.
It might then seem that (\ref{eq:atrial}) is assuming a \textit{minimal model}, where the number of intersections is precisely $\langle w_i, w_j \rangle$.
However, on closer inspection it turns out that this is not the case.
Whenever extra intersections between zig-zag paths are introduced, they always come in pairs of opposite orientations:
one where $\alpha_j$ is counter-clockwise from $\alpha_i$ and a second one vice-versa.
Expressions (\ref{eq:R_from_b1}), (\ref{eq:R_from_b2}) for $R$-charges can still be trusted in this case and we can see that the charges of the two extra fields are related (here assuming $i<j$):\footnote{
Note that this corresponds to a relation like $T_{12}+T_{34}=2$ in our analysis in Section~\ref{sec:UrbanRen}.}
\begin{equation}
	R_{ij} + R_{ji} = (b_j - b_i) + (2 + b_i - b_j) = 2 ~.
\end{equation}
This in turns shows that the $a$-trial remains unchanged:
\begin{equation}
	\Delta a
	= \frac{9}{32}\left( (R_{ij} - 1)^3 + (R_{ji} - 1)^3 \right)
	= \frac{9}{32}\left( (R_{ij} - 1)^3 + (2 - R_{ij} - 1)^3 \right)
	= 0 ~.
\end{equation}

In other words, the coefficient in front of $(R_{ij}-1)^3$ in (\ref{eq:atrial}) is not the total number of intersections between $\alpha_i$ and $\alpha_j$, but the sum of \textit{oriented intersections}, which is always $\langle w_i, w_j \rangle$.
Therefore (\ref{eq:atrial}) is valid for all, and not just for minimal models.

\subsection{Explicit $\tau_R$ and two-dimensional $a$-maximization}

Now that we have $b_i$ that maximize $a$-trial it is in fact possible to write down an explicit formula for $\tau_R$. Let us consider the rhombus lattice on the complex plane in Figure~\ref{fig:zigzags}. We take it to be periodically identified $z \sim z + c \sim z + c \tau_R$. Now if we follow a zig-zag path with winding numbers $(p,q)$ until we reach an identified point, the displacement on the complex plane is
\begin{equation}
\label{eq:Deltaz_pq}
	\Delta z_{(p,q)} = c(p + q \tau_R) ~.
\end{equation}
On the other hand, we can calculate the displacement from the parameters $b_i$. Take as an example the zig-zag path $\alpha_1$ in Figure~\ref{fig:zigzags}. As the zig-zag path crosses each rhombus, it intersects another path $\alpha_i$. As can be seen from the figure, the displacement after crossing the rhombus is given by the edge parallel to $\alpha_1$, which is just $e^{i\pi b_i}$, where $b_i$ is the parameter associated with the intersected path $\alpha_i$. More precisely the displacement is $\Delta z = - e^{i \pi b_i}$ if the intersection is positive (that is $\alpha_i$ is counter-clockwise from $\alpha_1$) and $\Delta z = e^{i \pi b_i}$ if negative. For example, one can see that the displacement is $e^{i\pi b_3}$ after intersection with $\alpha_3$ and it is $-e^{i \pi b_2}$ after intersection with $\alpha_2$. Therefore, the total displacement after going around a zig-zag path $(p,q)$ is
\begin{equation}
	\Delta z_{(p,q)=w} = - \sum_{i=1}^t \langle w, w_i \rangle  e^{i \pi b_i}
\end{equation}
because $\langle w, w_i \rangle$ is the number of positive intersections (minus the number of negative) with each path $\alpha_i$.

Now we can pick two zig-zag paths $(p_a,q_a)$ and $(p_b,q_b)$, apply (\ref{eq:Deltaz_pq}):
\begin{equation}
	\frac{\Delta z_{(p_a,q_a)}}{\Delta z_{(p_b,q_b)}} =
	\frac{p_a + q_a \tau_R}{p_b + q_b \tau_R}
	=
	\frac{\sum_i \langle w_a, w_i \rangle  e^{i \pi b_i}}{\sum_i \langle w_b, w_i \rangle  e^{i \pi b_i}} ~,
\end{equation}
and solve for $\tau_R$. The answer, of course, should not depend on which two paths we pick. We find indeed that it doesn't, and the result is a compact expression for $\tau_R$ in terms of the toric data and $b_i$:
\begin{equation}
\label{eq:tauR}
\boxed{
	\tau_R = - \frac{\sum_i p_i e^{i \pi b_i}}{\sum_i q_i e^{i \pi b_i}} ~.
}
\end{equation}
Note, by the way, this is just $\tau_R = \Delta z_{(0,1)} / \Delta z_{(1,0)}$, even though we cannot guarantee that paths $(0,1)$ and $(1,0)$ always exist.

According to~\cite{Butti:2005vn} the maximization of $a(b_i)$ in (\ref{eq:atrial}) over a $(t-1)$-dimensional space can be further reduced to a two-dimensional maximization of $a(x,y)$.
In this section we begin to examine the interplay of $(x,y)$ parameterization of the isoradial embeddings and $\tau_R$.

It turns out that for any point in the interior of the toric diagram with real coordinates $(x,y)$ we can assign trial $a_i$ charges as:
\begin{align}
\label{eq:aifromli}
	a_i(x,y) &= \frac{2 l_i(x,y) }{\sum_j l_j(x,y)} ~, \\
\label{eq:li}
	l_i(x,y) &\equiv \frac{\langle v_{i-1}, v_i \rangle}{\langle r_{i-1}, v_{i-1}\rangle \langle r_i, v_i \rangle} ~, \\
	r_i(x,y) &\equiv V_i - (x,y) ~, \\
	v_i &\equiv V_{i+1} - V_{i} ~.
\end{align}
Here $V_i$ are the coordinates of toric diagram's external points, $v_i$ are the vectors along external edges, and $r_i$ are vectors from the point $(x,y)$ to the external points.
If we now use these $a_i(x,y)$ to write the full $a$-trial function through (\ref{eq:b_from_a}) and (\ref{eq:atrial}), we get a function $a(x,y)$, which, when maximized gives the same $a_i$ as the full function $a(b_i)$.
For the details and the proof of this statement see~\cite{Butti:2005vn}.

An interesting observation is that at any point in the parameter space $(x,y)$, we can apply (\ref{eq:tauR}) to get a function $\tau_R(x,y)$ over the interior of the toric diagram.
Since $\tau_R$ also has two real parameters it is plausible that we can invert the function and in the end express $a$-trial purely as a function of $\tau_R$ and $\overline{\tau}_R$.
This would translate to the maximization of $a(\tau_R)$ over the space of complex structures which might have some nice geometric interpretation.
For now, let us content ourselves with the example of $\IC^3$.
From its details in Appendix~\ref{app:tauexamples}, we see that $\tau_R = \frac{-1 + e^{2 \pi i y}}{1 - e^{-2 \pi i x}}$, so that upon inverting we have
$e^{2 \pi i x} = \frac{\tau_R + \tau_R \overline{\tau}_R}{\overline{\tau_R} + \tau_R \overline{\tau}_R}$ and $e^{2 \pi i y} = \frac{1 + \tau_R}{1 + \overline{\tau}_R}$.
Whence, we have the desired formula of the trial $a$-function as
\be
a_{trial} = \frac{27}{4(2 \pi i)^3} \log\left(\frac{\tau_R + \tau_R \overline{\tau}_R}{\overline{\tau}_R + \tau_R \overline{\tau}_R} \right) \log\left(\frac{1 + \tau_R}{1 + \overline{\tau}_R} \right)( 2 \pi i - \log\left( \frac{\tau_R}{\overline{\tau}_R}\right)) ~,
\ee
to be maximized over the parameters $\tau_R$ and $\overline{\tau}_R$.

\section{$a$-maximization and invariants in terms of 
  generalized incidence matrix }
\label{sec:five}
\setall
Encouraged by the discovery that $\tau_R$ is invariant under Seiberg duality,
 a  quantity undiscussed in field theories prior to \cite{Jejjala:2010vb, hhjprr}, we now turn to the search for more invariants.
To this end, we will first examine $a$-maximization from a new perspective.

\subsection{Lagrange multipliers and $a$-maximization}
\label{amaxImproved}
We recall that the $a$-maximization procedure is the following constrained extremization problem for a dimer model with the sets of edges, faces, and vertices denoted, respectively, as $E,F,V$, we must maximize
\bea
a = \sum_{i \in E} (R_i-1)^3 ~,
\eea
subject to two conditions.
\begin{enumerate}
\item For each face $F$, we have
\bea\label{facecondition}
\sum_{i \in \pa F} ( 1 - R_i ) = 2 ~,
\eea
with the sum over edges that bound the face $F$.
The condition $i\in \pa F$ indicates that edge $i$ is in the boundary of face $F$.

\item For each vertex $V$ we have
\bea
\sum_{i\in \pa^{-1} V} R_i = 2 ~,
\eea
with sum over edges incident on $V$.
The condition $i\in \pa^{-1} V$ indicates that the vertex $V$ has the edge $i$ incident on it.
It is also useful to recast this as
\bea\label{vertexcondition}
\sum_{i\in \pa^{-1} V} (1 - R_i) = d_V - 2 ~,
\eea
where $d_V$ is defined as the valency of the vertex, \textit{i.e.}, the number of edges incident on the vertex, thus, $d_V := \sum_{i\in \pa^{-1} V} 1$.
\end{enumerate}

If we apply the usual technique of solving the linear constraints, then substituting back into the cubic and extremizing, we could have complicated polynomials.
We now use the Lagrange multiplier method which, as will be shortly seen, allows for some nice simplifications.
Define the standard function
\bea
A & = & \sum_{i=1}^{d} (R_i-1)^3 + \sum_{F} \lambda_F (2 - \sum_{i\in \pa F} (1 - R_i)) + \sum_{V} \lambda_V (2 - \sum_{i\in \pa^{-1} V} R_i) \cr
 & = & \sum_{i=1}^{d} (R_i-1)^3 + \sum_{F} \lambda_F (2 - \sum_{i\in \pa F} (1 - R_i)) + \sum_{V} \lambda_V ((d_V - 2) - \sum_{i\in \pa^{-1} V} (1 - R_i)) \nnm ~, \\
\eea
where we have explicitly marked the sum $i \in E$ by denoting the total number of edges as $d$.
The Lagrange multipliers for each $V$ and for each $F$ is denoted respectively as $\lambda_V$ and $\lambda_F$.

To extremize $A$, we need to set to zero the partial derivatives for each edge (field) $i$,
\bea\label{dRi}
\pa_{R_i} A = 3 (R_i-1)^2 + \sum_{F: i\in \pa F} \lambda_F + \sum_{V: \pa (i) = V} \lambda_V ~,
\eea
which will always give us quadratic equations for $R_i$, with the coefficients involving the multipliers $\lambda_V, \lambda_F$, which will be determined by solving the constraints at the end.
Each edge bounds two faces, call them $F_i^{+}, F_i^{-}$, and two vertices $V_i^+, V_i^-$.

A cleaner way to write the equations (\ref{dRi}) is to define a \textbf{generalized incidence matrix}, which we call $\cM$.
Usually, incidence matrices are defined for graphs, which have edges and vertices.
Here we have an \textit{embedded graph}, which has edges, faces, and vertices.
The matrix $\cM_{ij}$ has the row index $i$ running over the list of vertices and faces.
The index $j$ runs over the edges.\footnote{
Such structures have a history in the mathematics literature (see, for example,~\cite{Coxeter}).
We can define a rectangular matrix $\Pi^{(2)}$ whose columns are faces and whose rows are edges, and the entry $\Pi^{(2)}_{ji}$ counts the number of times the edge $j$ bounds the face $i$.
Similarly, we can define another rectangular matrix $\Pi^{(1)}$ whose columns are edges and whose rows are vertices, and the $\Pi^{(1)}_{ij}$ is zero or one depending on whether the vertex $i$ is an endpoint of the edge $j$.
The matrix $\cM$ we have constructed combines these as $\left( \begin{array}{c} \Pi^{(1)} \cr {}^t\Pi^{(2)} \end{array} \right)$.}
Since we have $|V|+|F|-|E| = 0$ from Euler's theorem applied to a genus one Riemann surface, this is remarkably a square matrix.
When $i$ is a vertex $V$, the entries of $\cM_{ij}$ are $1$ or $0$, depending on whether the edge $j\in \pa^{-1} V$ or not.
This part of $\cM$ is just the incidence matrix of the graph.
When $i$ is a face $F$, then the entry of $\cM_{ij}$ is $0$ if the edge $j$ does not bound the face, $1$ if the edge $j$ appears once as the boundary of the face is traversed, and $2$ if the edge appears twice (once from each side).
The last happens whenever $i$ transforms as an adjoint field under the gauge group corresponding to $F$ rather than as a fundamental or antifundamental.

There is another nice way to interpret the matrix $\cM$, namely, in terms of the associated rhombus lattice~\cite{Hanany:2005ss}.
In the mapping from dimer to rhombus lattice edges turn into faces (rhombi) while both faces and vertices turn into vertices.
The two adjacent faces and two adjacent vertices to a given edge in the dimer become the four corners of the rhombus.
Therefore $\cM_{ij}$ is $1$ precisely if rhombus $j$ has vertex $i$ as a corner and $0$ otherwise.
It can also be $2$ if the same vertex appears in two corners of the rhombus due to periodicity.

We can therefore write
\bea
\pa_{R_i } A = 3 (R_i - 1)^2 - \sum_{k\in \{V, F \}} \lambda_k \cM_{ki} ~.
\eea
Let us also define
\bea
\mu_i = ( 1 - R_i ) ~.
\eea
Then the condition $\pa_{R_i} A = 0$ becomes, rather succinctly,
\begin{equation}\label{defCon}
3 \mu_i^2 = \sum_{k\in \{V, F \}} \lambda_k \cM_{ki} ~.
\end{equation}

The linear constraints can also be written elegantly in terms of the incidence matrix $\cM$.
To do this we would like to package (\ref{facecondition}) and (\ref{vertexcondition}) into a single equation.
To this effect, define a vector $\cD_i$, where $i$ runs over faces and vertices.
When $i$ corresponds to a face, we put $\cD_i = 2$.
When $i$ corresponds to a vertex, we put $\cD_i = (d_i-2)$, where $d_i$ is the valency of the vertex.

Note that the vector $\cD_i$ carries no additional information beyond what is already in the matrix $\cM_{ij}$.
When $i$ corresponds to a vertex, the sum over edges $j$
\begin{equation}
\sum_{j=1}^d \cM_{ij} = d_i = \cD_i + 2 ~,
\end{equation}
as the total number of non-zero entries equals the number of edges that emanate from the vertex.
When $i$ corresponds to a face, the corresponding element of the vector $\cD_i$ is always $2$, and so conveys no new information.

Now, the two equations (\ref{facecondition}) and
(\ref{vertexcondition}) can be written as
\bea
\label{linconstsM}
\sum_{j\in E} \cM_{ij} \mu_j = \cD_{i} ~,
\eea
with the sum on $j$
running over all the edges, $1, \ldots, d$.
Typically the $d$ equations (\ref{linconstsM}) are not independent.
Rather, the number of independent equations is equal to the rank of the matrix:
\bea
\label{counteqs}
{\rm Rank} (\cM) = d - |{\rm Ker}(\cM) | ~,
\eea
where ${\rm Ker}(\cM)$ is the space of null
eigenvectors of $\cM$ given by the eigenvectors with zero
eigenvalues.

This suggests that the null eigenvectors of $\cM$ contain useful information about the $R$-charges.
Suppose $\cN^a_i$ is a null eigenvector, where $a$ runs over the possible null eigenvectors and $i$ runs over the $d$ components.
From the definition of null eigenvector, we have
\bea
\sum_{j} \cM_{ij} \cN^a_j = 0 ~.
\label{eq:nullvec}
\eea
Using this in (\ref{defCon}), we obtain
\bea
\sum_{i} \mu_i^2 \cN^a_{i} = \sum_{k} \lambda_k \cM_{ki} \cN^a_i = 0 ~.
\eea
In fact, if we consider the set of equations
\begin{equation}\label{keyeqs}
\boxed{
\begin{aligned}
\hspace{3 cm} \sum_{j=1}^d \cM_{ij} \mu_j & = \cD_{i} ~, \hspace{3cm} \\
\hspace{3 cm} \sum_{i=1}^d \mu_i^2 \cN^a_{i} & = 0 \hspace{3cm}
\end{aligned}
}
\end{equation}
we have precisely $d$ equations for $d$ variables, thus allowing us to find the $R$-charges via $R_i = \mu_i + 1$.
As argued in~\cite{Kato:2006vx} from the gravitational dual perspective, there is in fact a unique solution to the $a$-maximization problem.
In our case, the solution set of the boxed equations will give us the $R$-charges which maximize $a$ upon imposing the unitarity bound $R\geq \frac{2}{3}$ for gauge invariant operators.
In fact, for practical purposes, the weaker requirement $R> 0$ is enough to select the correct $R$-charge assignments.

\subsection{Properties of the generalized incidence matrix}
It is useful here to make the connection with rhombi and zig-zag paths again.
It is a well known fact~\cite{Hanany:2005ss} that the number of independent trial $R$-charges after solving the linear constraints is $(t-1)$ where $t$ is the number of external points in the toric diagram of the theory.
We can also see this explicitly from the zig-zag paths on the dimer.
Every consistent dimer has $t$ zig-zag paths and the remaining $(t-1)$ variables in the trial $a$-function can in fact be associated with the $(t-1)$ paths which are linearly independent (in the appropriate sense mentioned below).
It is therefore nice to see these zig-zag paths appear naturally here:
they are precisely the null vectors $\cN^a_i$ of matrix $\cM$!

Let us elaborate on what this means.
First, we assign a $d$-dimensional vector to a zig-zag path $\alpha_a$ in the dimer as follows:
\begin{equation}
\label{eq:cN_from_zigzag}
	\cN^a_i =
	\begin{cases}
 	 +1 & \text{if $\alpha_a$ intersects edge $i$ after turning right at a node } \\
 	 -1 & \text{if $\alpha_a$ intersects edge $i$ after turning left at a node } \\
 	0 & \text{if $\alpha_a$ does not intersect edge $i$ }
 \end{cases} ~.
\end{equation}
The overall sign does not matter.
The point is that edges are included with alternating signs.
This vector, in fact, represents the charge assignments of the baryonic $U(1)$ symmetry associated with $\alpha_a$.

We can prove that such $\cN^a_i$ is a null vector of $\cM$.
Consider $\sum_{j} \cM_{ij} \cN^a_j$ for a fixed $i$ which is now either a face or a vertex.
Going to the rhombus lattice picture, the zig-zag path is a sequence of rhombi with alternating signs (otherwise known as ``rhombus path'' or ``train track''), while $i$ is a vertex.
The sum simply counts how many times the vertex $i$ is included in positive faces minus the negative faces, but since each vertex is included in precisely two adjacent faces, the sum will be zero for all $i$ (see Figure~8 in~\cite{Hanany:2005ss}).
Therefore, $\cN^a_i$ \textit{is} a null vector.
Now, to show that vectors $\cN^a_i$ in fact cover the whole ${\rm Ker}(\cM)$ we simply cite the fact that there are $(t-1)$ independent $\cN^a_i$ among the $t$ zig-zag paths, corresponding to $(t-1)$ anomaly free $U(1)$ symmetries.
A single relationship between them is that $\sum_{a=1}^t \cN^a_i = 0$, because each edge is included twice with opposite signs.
In the end if we take $\cN^a_i$ according to (\ref{eq:cN_from_zigzag}) for any chosen $(t-1)$ zig-zag paths we get a basis for the full ${\rm Ker}(\cM)$.

In conclusion, our generalized adjacency matrix has the property that
\begin{equation}
\fbox{$	|{\rm Ker}(\cM) | = t - 1 $}\ ,
\end{equation}
where $t$ is the number of external points in the toric diagram.
An immediate corollary of this is that $|{\rm Ker}(\cM)|$, a rough measure of the complexity of the $R$-charges as it is the number of quadratic equations, is preserved under Seiberg (toric) duality since it depends only on the toric diagram.
One can easily verify this and for reference, we tabulate the results in Table~\ref{t:S}.
\begin{table}[ht!]
\[
\begin{array}{|c|c|}\hline
{\rm CY}_3 \mbox{ geometry } & |{\rm Ker}(\cM)| \\ \hline \hline
\IF_0 & 3 \\ \hline
L^{2,2,2} & 5 \\ \hline
L^{2,3,2} & 6 \\ \hline
L^{2,4,2} & 7 \\ \hline
L^{3,3,3} & 7 \\ \hline
dP_2 & 4 \\ \hline
dP_3 & 5 \\ \hline
PdP_3b & 5 \\ \hline
\end{array}
\]
\caption{{\sf \small
The dimension of the null space of the matrix $\cM$ is the same for all the Seiberg dual phases of a dimer theory.
}}
\label{t:S}
\end{table}

Another visual and interesting consequence from $|{\rm Ker}(\cM)|$ can be obtained by summing over $i$ in (\ref{eq:nullvec}).
Indeed, it happens that $\sum_{i=1}^d \cM_{ij} = 4$ because a fixed edge $j$ must connect two vertices, a black and a white node from the bipartite condition, and as well separates two faces (which might be the same).
As $\sum_i\,\cM_{ij}$ is independent on $j$, it then follows that $\sum_j\,\cN^a_j=0$ for all $a$.
Thus, we might think of them as defining the gauged linear sigma model (GLSM) charges for a Calabi--Yau variety.
The role of this variety, which can be seen experimentally not to be invariant under Seiberg duality, is not clear.

Summing the first equation in~\eref{keyeqs} over $i$, we may conclude that
\begin{equation}
\sum_{i,j=1}^d \cM_{ij} \mu_j = \sum_{i=1}^d \cD_i = 4|F| ~.
\end{equation}
Since $\mu_i = 1 - R_i$ and the $R$-charges must be real and positive, on physical grounds, we demand that $\mu_i < 1$ for all $i$.
To see~\eqref{keyeqs} in action, let us apply to a few of our familiar examples.

\subsection{Applications: $\IC^3$ and SPP}
\paragraph{$\IC^3$ calculation:}
In this case
\bea
\cM = \begin{pmatrix} 1 & 1 & 1 \\ 1 & 1 & 1 \\ 2 & 2 & 2 \end{pmatrix} ~, \quad
\cD = \begin{pmatrix} 1 \\ 1 \\ 2 \end{pmatrix} ~.
\eea
The first row corresponds to the black vertex, the second to the white vertex, the third to the face.
The $2$ corresponds to the fact that as the boundary of the face is traversed, we encounter each edge twice; that is, all three fields here are adjoints.

In this case two linearly independent null vectors are
\bea
\cN^{(1)} = \begin{pmatrix} 1 \\ 0 \\ -1 \end{pmatrix} ~, \quad
\cN^{(2)} = \begin{pmatrix} 0 \\ 1 \\ -1 \end{pmatrix} ~.
\label{eq:c3null}
\eea

Using these null vectors, the second of (\ref{keyeqs}) leads to equations
\bea \label{c3-m}
\mu_1^2 = \mu_2^2 = \mu_3^2 ~.
\eea
Next, the first equation in (\ref{keyeqs}) gives
\bea \label{c3-lin}
\mu_1 + \mu_2 + \mu_3 = 1 ~.
\eea
Hence, combining \eqref{c3-m} with \eqref{c3-lin} and demanding that the $R$-charges be positive, we retrieve the familiar result that for all $i=1,2,3$,
\begin{equation}\label{$R$-c3}
\mu_i = {1 \over 3} ~, \qquad R_i = {2 \over 3} ~.
\end{equation}

Note that, as promised, the sum of the entries of each $\cN$ vector is zero.
This ensures that when we take the quotient $\IC^3 // \cN^{a,\,\dagger}$, we have a Calabi--Yau space.
As we have stated, the null vectors can be thought of as the set of charges defining the GLSM for the resulting Calabi--Yau variety.
Taking ${\rm Ker}(\cN^{a,\,\dagger})$, this is one-dimensional and therefore $\mathbb{C}$.
We can understand this as follows:
$\cN^{(1)}$ yields an identification $z_1 \sim -z_2$ while $\cN^{(2)}$ yields an identification $z_2 \sim -z_3$.
Thus, we are left with one coordinate.

\paragraph{SPP calculation:}
Next, let us address the suspended pinched point (SPP), another well known theory.
In this case the number of edges is $d = 7$.
The dimer, generalized incidence matrix $\cM$, and the $\cD$ vector are, respectively,
\bea
\begin{array}{l}
\resizebox{!}{5cm}{\includegraphics{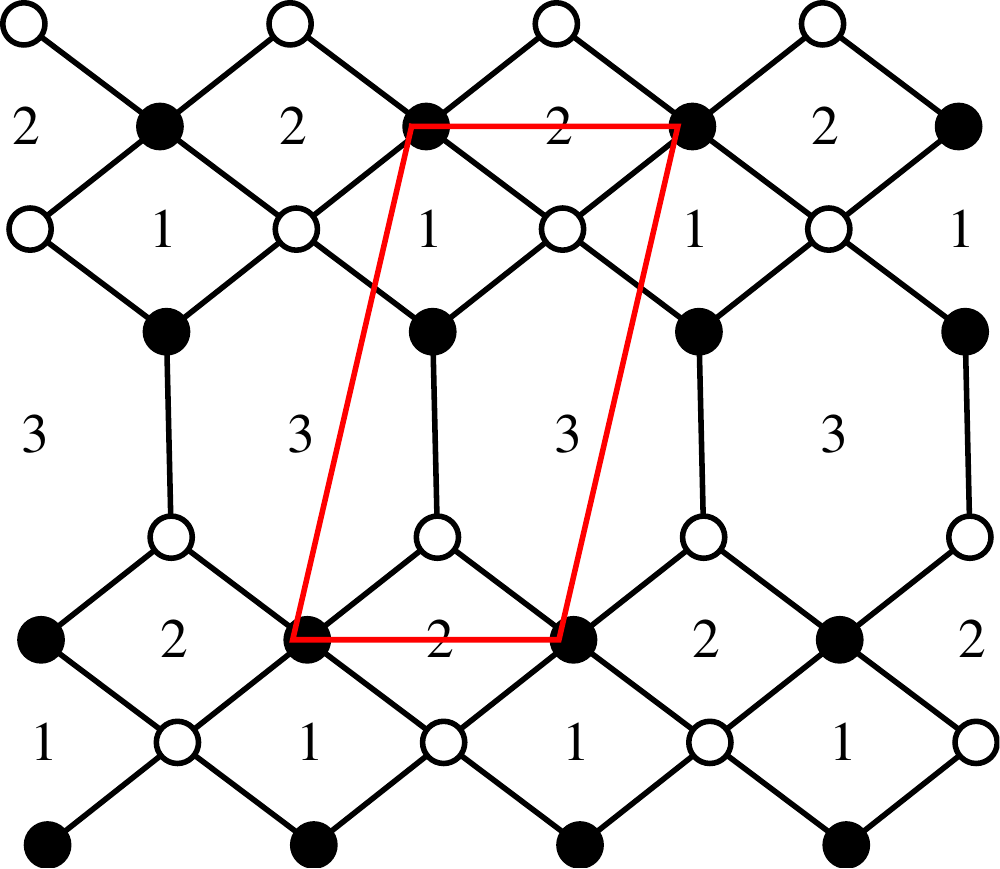}}
\end{array}
\cM = \begin{pmatrix}
0 & 0 & 1 & 1 & 0 & 0 & 1 \\
1 & 1 & 0 & 0 & 1 & 1 & 0 \\
0 & 0 & 0 & 0 & 1 & 1 & 1 \\
1 & 1 & 1 & 1 & 0 & 0 & 0 \\
1 & 1 & 1 & 1 & 0 & 0 & 0 \\
1 & 1 & 0 & 0 & 1 & 1 & 0 \\
0 & 0 & 1 & 1 & 1 & 1 & 2 \\
\end{pmatrix} ~,~
\cD = \begin{pmatrix} 1 \\ 2 \\ 1 \\ 2 \\ 2 \\ 2 \\ 2 \end{pmatrix} ~.
\eea
The first row corresponds to the first black vertex, which is trivalent and so has three entries equal to $1$.
The second row is the second black vertex of valency four and four entries of $1$ at the corresponding edges.
The third and fourth rows correspond to the white vertices.
The fifth, sixth, and seventh rows correspond to the three faces.
The columns correspond to the fields $\{ X_{12}, X_{21}, X_{13}, X_{31}, X_{23}, X_{32}, X_{33} \}$.

There are four null vectors:
\bea
\{\cN^{(1)}, \cN^{(2)}, \cN^{(3)}, \cN^{(4)} \} =
\begin{pmatrix} 1 \\ -1 \\ 0 \\ 0 \\ 0 \\ 0 \\ 0 \end{pmatrix} ~,~
\begin{pmatrix} 0 \\ 0 \\ 1 \\ -1 \\ 0 \\ 0 \\ 0 \end{pmatrix} ~,~
\begin{pmatrix} 0 \\ 0 \\ 0 \\ 0 \\ 1 \\ -1 \\ 0 \end{pmatrix} ~,~
\begin{pmatrix} 1 \\ 0 \\ -1 \\ 0 \\ -1 \\ 0 \\ 1 \end{pmatrix} ~,~
\label{eq:sppnull}
\eea
corresponding to the fact that SPP has $t=5$ external vertices in the toric diagram

Note that just as in~\eref{eq:c3null}, the null vectors in~\eref{eq:sppnull} have elements that sum to zero.
In this case the corresponding Calabi--Yau is a conifold.
(The first three null vectors reduce $\IC^7$ to $\IC^4$ and the fourth relates the coordinates of $\IC^4$ by the conifold relation.)

The first three null vectors offer the equations
\begin{equation}
\mu_{12}^2 = \mu_{21}^2 ~, \quad
\mu_{13}^2 = \mu_{31}^2 ~, \quad
\mu_{23}^2 = \mu_{32}^2 ~.
\end{equation}
The last null vector though does not give a linear equation, but, rather, a quadratic:
\begin{equation}\label{compNul}
\mu_{12}^2 - \mu_{13}^2 - \mu_{23}^2 + \mu_{33}^2 = 0 ~.
\end{equation}
From the $\cM \cdot \mu = \cD $ we obtain three equations:
\bea
&& \mu_{12} + \mu_{21} - \mu_{33} = 1 ~, \cr
&& \mu_{13} + \mu_{31} - \mu_{33} = 1 ~, \cr
&& \mu_{23} + \mu_{32} + \mu_{33} = 1 ~.
\eea
Solving this system of equations, there are ten sets of solutions for the $\mu_{ij}$, of which only one has all of the $R$-charges being positive.
This reproduces the known charges of the fields of SPP:
\bea \label{$R$-SPP}
&& R_{12} = R_{21} = 1 - {1 \over \sqrt{3}} ~, \cr
&& R_{13} = R_{31} = R_{23} = R_{32} = {1 \over \sqrt{3}} ~, \cr
&& R_{33} = 2 - {2 \over \sqrt{3}} ~.
\eea

We have applied this technique to all the consistent theories in the classification of~\cite{Davey:2009bp} and find complete agreement with the the traditional methods.
This shows that the solution of the $R$-charges can be simply expressed in terms of the generalized incidence matrix in terms of the two equations (\ref{keyeqs}).
The simple consideration in (\ref{counteqs}) suggests that the number of independent equations in (\ref{keyeqs}) is exactly the right number to give all the $R$-charges.

\section{Number theoretic invariants of Seiberg duality}
\label{sec:six}

We have established in Section~\ref{sec:UrbanRen} that $\tau_R$ is an invariant under the urban renewal operation that implements Seiberg duality on a dimer.
The \textit{master space}, which is the solution to the F-term equations of an ${\cal N}=1$ supersymmetric gauge theory, matches the IR moduli space of a quiver gauge theory~\cite{Forcella:2008bb}.
Its dimension is $F+2$, where $F$ is the number of $U(N)$ gauge groups, \textit{i.e.}, the number of faces in the dimer.
Generically, this master space is reducible and separates into pieces of different dimensionality.
The maximal dimensional piece of the variety ---
the \textit{coherent component} or \textit{equidimensional hull} under a primary decomposition ---
is an $(F+2)$-dimensional Calabi--Yau cone.
The dimension of the coherent component is also invariant under Seiberg duality~\cite{Forcella:2008ng}.
Motivated by our discussion of $R$-charges, in this section we scan for further new invariants of Seiberg duality.

It has been observed before that the $R$-charges determined by $a$-maximization are algebraic numbers.
These are numbers which are roots of polynomials with coefficients that are rational, \textit{i.e.}, belonging to $\mQ$.
The collection of all algebraic numbers forms $\bmQ$, a field closed under addition and multiplication, and obeying the relevant axioms.

Since the $R$-charges after Seiberg duality are obtained simply by leaving a subset of the $R$-charges fixed, and using \textit{field operations} of addition using another subset, this means that the field of definition of the $R$-charges, and in particular the degree of that extension is unchanged.
So we have additional number theoretic invariants under Seiberg duality (along with $j(\tau_R)$).

Indeed, the $R$-charges of our theories are algebraic numbers, arising from systems of polynomials, during $a$-maximization, in integer coefficients.
The gravity dual statement that the volume of the Sasaki--Einstein manifolds are algebraic is discussed in~\cite{Martelli:2005tp}.

It is an interesting question to ask in what precise number field do the $R$-charges live.
After all, Grothendieck's original intent in studying the dessin d'enfants, which are an equivalent description of our dimer models~\cite{Jejjala:2010vb, hhjprr}, was to investigate the absolute Galois group~\cite{grot}.
The absolute Galois group ${\rm Gal}(\bmQ/\mQ)$ acts as a symmetry of the field of algebraic numbers, which preserves the rational numbers.
Moreover, it acts faithfully on the set of dessins.

For $\IC^3$, we have that the $R$-charges, given in \eqref{$R$-c3}, are still in $\IQ$.
For SPP, the $R$-charges, given in \eqref{$R$-SPP}, are in $\IQ[\sqrt{3}]$, the rationals adjoining a single radical, \textit{viz.}, $\sqrt{3}$.
Recalling that the \textit{degree} of a field extension $K$ over $L$, commonly denoted as $[K:L]$, is the dimension of the vector space when writing elements of $K$ as vectors with entries in $L$ and with basis as the numbers introduced in the extension, the degree of the minimal field extension for SPP turns out to be two.

It is instructive to tabulate the minimal number fields over the rationals wherein the $R$-charges reside, for the 42 consistent dimers/tilings catalogued in~\cite{Davey:2009bp}.
In Table~\ref{t:R}, we list the degree of extension, starting from the trivial degree one, when the $R$-charges are rational.\footnote{
In the last entry in Table~\ref{t:R}, $f(x)$ is:
{\Tiny
\bea
&& 129895959052919063627088658824484033170106970563057053464097738113158357486077961163576360759355431959070613052958136429700{\rm -} \nn \\
&& 7990706677245120266223000216858808331489576634388895611486208 + 812780867079221751989485691592256982562445490116486994857815{\rm -} \nn \\
&& 32406074736034254990618952381740902301999847184924691931671841243521462114222422531961018099657867264 x - 1605590251086677055{\rm -} \nn \\
&& 032517976888916263032528341347296581174611981547691619152719002618787800157726841563671420301829401419343371688614035456 x^2 - \nn \\
&& 58501895038049041480240324513991980662242406126145870387880078437110541099265053803105661755970304429347559234666496 x^3 + \nn \\
&& 694785402847564626838097983555518880082522398594151133211393984791686345118886967099469533184 x^4 + \nn \\
&& 8748375862087189421259901955556860436477229485948805897674628434108672 x^5 - 68749725604677082362249064632675381493544491440 x^6 - \nn \\
&& 162855999849603150949559 x^7 + x^8 ~, \nn \eea }
where we have hyphenated the integer coefficients, due to their lengths, at orders zero, one, and two in $x$.}
For completeness, we shall use the pair notation (theory, minimal field containing the $R$-charges).
We point out that $L^{1,1,1}$ is
commonly known as the conifold, $L^{1,2,1}$ as SPP, and
$\IC^3/\IZ_3$ as $dP_0$.

\begin{table}[ht!]
\[
\begin{array}{|c|c|}\hline
\mbox{Extension Degree} & \mbox{Theory} \\ \hline \hline
1 &
\begin{array}{l}
(\IC^3, \IQ) \ ,
(\IF_0, \IQ) \ ,
(dP_3, \IQ) \ ,
(Y^{3,0}, \IQ) \ ,
(Y^{4,0}, \IQ) \ , (Z^{3,1}, \IQ) \ ,
(L^{1,1,1}, \IQ)
\end{array}
\\ \hline
2 &
\begin{array}{l}
(L^{1,2,1}, \IQ[\sqrt{3}]) \ , (L^{1,3,1}, \IQ[\sqrt{7}]) \ ,
(L^{1,4,1}, \IQ[\sqrt{13}]) \ , \\
(L^{2,3,2}, \IQ[\sqrt{7}]) \ ,
(L^{2,4,2}, \IQ[\sqrt{3}]) \ ,
(L^{3,4,3}, \IQ[\sqrt{13}])\ , \\
(dP_1, \IQ[\sqrt{13}]) \ , (dP_2, \IQ[\sqrt{33}]) \ , \\
(PdP_3b, \IQ[\sqrt{5}]) \ , (PdP_3c, \IQ[\sqrt{3}]) \ , \\
(Y^{3,1}, \IQ[\sqrt{33}]) \ ,
\end{array}
\\ \hline
3 &
\begin{array}{c}
(PdP_4, \IQ[x]) : -3356856 + 128412 x - 1250 x^2 + x^3 =0 \,
\end{array}
\\ \hline
4 &
\begin{array}{c}
 ( PdP_2, \IQ [ x ] ) : -22719338592 + 343430020 x - 1613592 x^2
+ 2164 x^3 + x^4 =0
\end{array}
\\ \hline
8 & ( X^{3,0}, \IQ [ x ] ) : f(x) = 0
\\ \hline
\end{array}
\]
\caption{{\sf \small
The minimal field extensions of $\IQ$ containing the values of the $R$-charges.
We catalogue by the degree of the extension and include the actual field together with the theory in pairs.
The definition of $f(x)$ is given in the footnote to the main text.
Furthermore, we omit orbifoldings of these theories, as their field of extension is trivially identical to the parent theory.}}
\label{t:R}
\end{table}

Table~\ref{t:R} shows that different theories in fact have the same field extensions.
Note, for instance, that at degree one, we have $\IC^3$, the conifold, various orbifolds of these theories and $Z^{3,1}$, which is not related to the others.
At degree two, we have another orbifold example: $L^{2,4,2}$ has the same $R$-charges as the parent $L^{1,2,1}$ and thus the same field extension.
That $L^{1,3,1}$ has the same field extension as $L^{2,3,2}$ is more interesting.
In general, $L^{a,b,a}$ and $L^{d,b,d}$ with $d=b-a$ have identical fields extensions.
Indeed, this can be easily seen from the generic expressions for the $R$-charges (see, \textit{e.g.}, (149) in~\cite{Jejjala:2010vb}).
The $R$-charges for $L^{a,b,a}$ are rational numbers extended with $\sqrt{a^2+b^2-a\,b}$.
Hence, upon setting $a=b-d$, we find that this becomes $\sqrt{d^2+b^2-d\,b}$, which is the irrational factor extending $L^{d,b,d}$, and so the field extension is the same.
Incidentally, we note that $dP_1$ has the same field extension as $L^{1,4,1}$ and $L^{3,4,3}$ and that $dP_2$ has the same field extension as $Y^{3,1}$.

For most of the theories, we can easily read off the relevant extension from the list of charges.
For theories such as $PdP_2$, $X^{3,0}$, and $PdP_4$ the different $R$-charges involves roots of different polynomials.
For a set of algebraic numbers, there is polynomial $f(x) $ 
for the  generator $x$ of a unique field which contains the whole  set 
of algebraic numbers. Each algebraic number can be expressed as a  
sum of powers of $x$.\footnote{These computations are conveniently 
available through number theoretic commands in mathematical software,
 such as {\tt ToNumberField} in {\tt Mathematica}.}
 The defining polynomial $f(x)$ for $X^{3,0}$, is given in the previous 
footnote.
The existence of an $x$ for any set of 
algebraic numbers is guaranteed by the \textbf{primitive element theorem}.
A rudimentary discussion of this statement is given 
in  Appendix~\ref{app:alg}.

The  expression for each $R$-charge as a polynomial in $x$
 defines a vector, and the number of linearly independent vectors (that is, the rank of the subspace they span) is an invariant under Seiberg duality.
This can be understood recalling that starting with a given phase, the $R$-charges in the Seiberg dual phase are given in terms of field operations on the old ones.
If there are no massive fields to integrate out, this implies that the new $R$-charges, expressed as vectors in the form we described, are linear combinations of the old vectors, and so the rank of the subspace is the same.
Suppose there is a mass term to integrate out.
Then we have, say, $R_1 + R_2 = 2$ from a two-valent node, and $R_1 + \sum_i X_i = 2$, $R_2 + \sum_j Y_j = 2$, the latter two conditions arising from other terms in the superpotential (\textit{i.e.}, the other nodes on which the fields with $R$-charges $R_1$ and $R_2$ are incident).
Naively, we might think that integrating out the mass term makes $R_1$ and $R_2$ disappear, and this would decrease the dimensionality of the vector space of $R$-charges.
But, as $R_1$ is in the vector space spanned by the $X_i$ and $R_2$ is in the vector space spanned by the $Y_j$, the dimension of the vector space of the $R$-charges does not change.
Thus, we have established that the rank of the subspace spanned by the vectors associated to $R$-charges in the prescribed way is a Seiberg duality invariant.

It is a curious fact that for nearly all the cases in Table~\ref{t:R}, the extension degree equals the number of linearly independent vectors.
The lone exception to this is at extension degree $8$, where the theory $X^{3,0}$ has $R$-charges that are vectors in a five-dimensional space.

\section{Concluding remarks and open questions}\setall
\label{sec:conc}

Let us summarize the main results.
We have examined a class of ${\cal N}=1$ four-dimensional quantum field theories arising as the worldvolume gauge theories on D$3$-branes at the tip of toric Calabi--Yau cones over a five-dimensional Sasaki--Einstein base.
This class of theories is distinguished by the fact that each field appears exactly twice in the superpotential $W$, once with a plus sign and once with a minus sign.
As a consequence, a bipartite graph (equivalently, a dimer model or brane tiling) depicts the field content of the theory and the superpotential interactions and encapsulates the brane realization of the theory.
Fixing the $R$-charges by $a$-maximization, a particular $R$-dimer is distinguished with a complex structure parameter $\tau_R$.

In this paper, we have proved that this $\tau_R$ is, up to $SL(2,\mathbb{Z})$ equivalence, an invariant under Seiberg duality, an operation which maps one dimer model to another. In Section~\ref{sec:UrbanRen} we showed this by 
using the known description of Seiberg duality as a composition of 
the urban renewal  move along with integrating out massive fields. 
We showed that each of these steps could be expressed as local moves 
on the dimer with shape determined by physical $R$-charges compatible with the 
isoradial embedding. In this process, the positions of the vertices do not change, hence the description as  a rigid local refinement.
In Section~\ref{sec:taufromtoric} we derived a formula for $\tau_R $ 
in terms of toric data of the Calabi--Yau. 

 We developed  a new expression of the $a$-maximization algorithm, 
in terms of a generalized incidence matrix, presented in 
Section~\ref{sec:five}. 
The dimension of the  kernel of this matrix was shown to 
be a  Seiberg duality invariant, in fact related to the 
number of zig-zag paths, which is  one less than the 
number of external points of the toric diagram. 

The generalized incidence matrix deserves further scrutiny.
This structure can, of course, be defined in higher dimensions for a cell complex (or simplicial complex), labeling rows with $0$-cells, $2$-cells, etc.\ and columns with $1$-cells, $3$-cells, etc.
Again for zero Euler character, this matrix is a square.
A venerable literature exists on this subject in mathematics~\cite{Coxeter}, and it would be interesting to elaborate on these ideas in the context of toric theories.

Pursuing the theme of number theoretic aspects of the physics
of toric CFTs, motivated by the dimer/dessins connections 
investigated in \cite{Jejjala:2010vb,hhjprr},  we considered the 
fields of definition of the set of $R$-charges of a given theory.  
The $R$-charges are in a minimal  extension field $\mathbb{Q}[x]$,
where $x$ is some algebraic number adjoined to the rationals.
The degree of the field extension is a number theoretic invariant 
of Seiberg duality. The set of $R$-charges forms  a vector space 
over $ \mathbb{Q} $, with basis associated with powers of $x$. 
The rank of this vector space was also shown to be Seiberg duality 
invariant.

Other investigations of connections between number theory, gauge theories, and Calabi--Yau have been conducted recently.
Calabi--Yaus over finite fields, in connection with mirror symmetry, have been studied in~\cite{candelas}.
The investigation of placing the CY$_3$ itself, as a gauge theory moduli space, on different number fields, is the subject of~\cite{He:2010mh}.
Hints of the string worldsheet as an object defined on algebraic number fields appear in the context of matrix models and topological strings over $\mP^1$~\cite{matxbelyi,gopakumar}.
Articulating a unifying picture of these diverse, but somewhat complementary, appearances of number theory in string physics, is an interesting problem.
Graph-theoretic moves for generating classes of   dimer models 
have  recently  investigated in~\cite{deMedeiros:2010pr, deMedeiros:2011ce}. 
The interplay of these methods with the combinatoric aspects of Dessins 
described in \cite{Jejjala:2010vb} may provide new  insights.

The invariance of  $\tau_R$ under Seiberg duality and its explicit 
expression in terms of toric data shows that it is a geometric property of 
the Calabi Yau, independent of the gauge theory used to to realize the CY 
as a moduli space. This suggests that this quantity has a deeper physical meaning in terms of branes. 
We can think of a brane tiling in terms of D$5$-branes associated to each of the faces and NS$5$-branes that wrap a holomorphic curve associated to the edges~\cite{Franco:2005rj}.
The toric diagram corresponding to the dimer model captures the alignment of $(p,q)$-five branes that recapitulates the quantum field theory in question~\cite{AharonyBH}.
The $(p,q)$-web thus obtained is the graph dual of the toric diagram.
The Newton polynomial defined by the toric diagram offers a thickening of the $(p,q)$-web and also specifies the mirror geometry~\cite{HoriKT}.
The CY$_3$ admits a special Lagrangian fibration in which the edges of the web provide the zero locus for a $(p,q)$ cycle of ${\mathbb T}^2$~\cite{FengGW}.
Understanding the relationship between these structures and the $\tau_R$ fixed by $R$-charges is the subject of a forthcoming paper~\cite{SLAGS}.

\section*{Acknowledgements}

We are delighted to thank Robert de Mello Koch, Rak-Kyeong Seong, and David Turton for discussions.
AH would like to thank the Isaac Newton Institute and Oxford University for kind hospitality during the completion of this project.
As well, AH and YHH thank the Mathematisches Forschungsinstitut Oberwolfach for generous hospitality. 
YHH is indebted to the gracious patronage of the Science and Technology Facilities Council, UK, for an Advanced Fellowship, the Chinese Ministry of Education, for a Chang-Jiang Chair Professorship at NanKai University, as well as City University, London and Merton College, Oxford, for their enduring support, and raises his glass and his soul to Elizabeth Katherine, for their impending nuptials.
YHH and VJ acknowledge NSF grant CCF-1048082.
VJ and SR are supported by an STFC grant ST/G000565/1.
SR thanks the Galileo Galilei Institute for hospitality during the final stages of the paper.
DRG thanks Benasque Center for Science for warm hospitality while this work was being finished.
DRG is supported by the Israel Science Foundation through grant 392/09.
He also acknowledges support from the Spanish Ministry of Science through the research grant FPA2009-07122 and Spanish Consolider-Ingenio 2010 Programme CPAN (CSD2007-00042).
Finally, YHH's coauthors toast the bride and the groom and wish them joy, good fortune, and connubial bliss.

\appendix

\section{Examples of $\tau_R$}
\label{app:tauexamples}

Here we provide examples of $\tau_R$ calculation directly from toric diagram using the method of Section~\ref{sec:taufromtoric}.

\subsection{$\IC^3$}

Let's pick the toric diagram corners as
\begin{equation}
	V_1 = (0,0), \quad V_2 = (1,0), \quad V_3 = (0,1) ~.
\end{equation}
The $r_i$ and $v_i$ vectors:
\begin{equation}
\begin{split}
	v_1 = (1, 0),& \quad 	r_1 = (-x, -y) ~, \\
	v_2 = (-1, 1),& \quad r_2 = (1-x,-y) ~, \\
	v_3 = (0, -1),& \quad r_3 = (-x,1-y) ~,
\end{split}
\end{equation}
and the primitive normals
\begin{equation}
	w_1 = (0, -1), \quad w_2 = (1, 1), \quad w_3 = (-1, 0) ~.
\end{equation}
Using the formulas (\ref{eq:aifromli}), (\ref{eq:li}):
\begin{equation}
	l_1 = \frac{1}{xy}, \quad l_2 = \frac{1}{y(1-x-y)}, \quad l_3 = \frac{1}{x(1-x-y)} ~,
\end{equation}
\begin{equation}
	a_1 = 2(1-x-y), \quad a_2 = 2x, \quad a_3 = 2y ~.
\end{equation}
Plugging these expressions in (\ref{eq:atrial}) we find the $a$-trial function:
\begin{equation}
	a(x,y) = \frac{27}{4}\,x\,y\,(1-x-y) ~.
\end{equation}
It has a maximum at
\begin{equation}
	\bar{x}=\bar{y}=\frac{1}{3}, \quad \bar{a}_1=\bar{a}_2=\bar{a}_3=\frac{2}{3}, \quad a(\bar{x},\bar{y}) = \frac{1}{4} ~.
\end{equation}
The $b_i$ parameters for primitive normals are
\begin{equation}
	b_1 = 2(1-x-y), \quad b_2 = 2(1-y), \quad b_3 = 2
\end{equation}
leading to the expression for $\tau_R$ according to (\ref{eq:tauR}):
\begin{equation}
	\tau_R(x,y) = \frac{-1 + e^{2\pi i y}}{1 - e^{-2\pi i x}} ~.
\end{equation}
At the maximum of $a$ this becomes
\begin{equation}
	\tau_R = e^{2 \pi i / 3} ~.
\end{equation}

\subsection{SPP}

Let us now analyze the simplest theory with irrational charges: suspended pinch point (SPP).
Taking the toric diagram as in Figure~\ref{fig:spptoric} we have:
\begin{equation}
	l_1 = \frac{1}{(2-x)(1-y)} ~, \quad
	l_2 = \frac{1}{(1-y)(x-y)} ~, \quad
	l_3 = \frac{1}{(x-y)y} ~, \quad
	l_4 = 0 ~, \quad
	l_5 = \frac{1}{y(2-x)} ~,
\end{equation}
and
\begin{equation}
	a_1 = \frac{2(x-y)y}{2-y} ~, \quad
	a_2 = \frac{2y(2-x)}{2-y} ~, \quad
	a_3 = \frac{2(2-x)(1-y)}{2-y} ~, \quad
	a_4 = 0 ~, \quad
	a_5 = \frac{2(1-y)(x-y)}{2-y} ~.
\end{equation}
This leads to the $a$-trial function
\begin{equation}
	a(x,y) = \frac{27}{4}\frac{(2-x)(x-y)(1-y)y}{2-y} ~,
\end{equation}
with the maximum at
\begin{equation}
	\bar{x} = \frac{9-\sqrt{3}}{6}, \quad \bar{y} = \frac{3-\sqrt{3}}{3}, \quad a(\bar{x},\bar{y}) = \frac{3\sqrt{3}}{8} ~,
\end{equation}
\begin{equation}
	\bar{a}_1= \bar{a}_2 = 1 - \frac{1}{\sqrt{3}} ~, \quad
	\bar{a}_3= \bar{a}_5 = \frac{1}{\sqrt{3}} ~, \quad
	\bar{a}_4 = 0 ~.
\end{equation}
The $\tau_R$ at an arbitrary point can be expressed as
\begin{equation}
	\tau_R(x,y) = \frac{
		\exp\left( 2\pi i \, \frac{x+y-xy}{2-y} \right)
		- \exp\left( 2\pi i \, \frac{(x-y)(1-y)}{2-y} \right)
	}{
		-2
		+ \exp\left( 2\pi i \, \frac{x+y-xy}{2-y} \right)
		+ \exp\left( 2\pi i \, \frac{x-y}{2-y} \right)
	} ~,
\end{equation}
which at the maximum of $a$ is
\begin{equation}
	\tau_R =
	\frac{\exp\left(2\pi i / \sqrt{3}\right) - 1}
	{3\exp\left(\pi i / \sqrt{3}\right) - 1}
	\simeq
	0.16465 + 0.539718 \, i ~,
\end{equation}
\begin{equation}
	j(\tau_R) \simeq -35175 ~.
\end{equation}
This matches the result of~\cite{Jejjala:2010vb}.

\section{Field extensions and the primitive element theorem}
\label{app:alg}
In this appendix, we use an illustrative example to explain some results in number theory which are used in the computation of the algebraic extensions over $\IQ$ for the $R$-charges.
Consider for example, the numbers $\sqrt{2}$ and $\sqrt{3}$.
One can show that these can be expressed in terms of a root of the polynomial
\bea\label{def23}
1 - 10 x^2 + x^4 = 0 ~.
\eea
Indeed, one checks that
\bea
&& (-{9\over 2} x + {1 \over 2} x^3)^2 = 2 ~, \cr
&& ({11 \over 2} x - {1 \over 2} x^3)^2 = 3 ~,
\eea
when the relation in (\ref{def23}) is used.
This means that we can write
\bea\label{expsalg}
&& \sqrt{2} = -{9\over 2 } x + {1 \over 2} x^3 ~, \cr
&& \sqrt{3} = {11 \over 2 } x - {1 \over 2} x^3 ~,
\eea
where $x$ is understood to satisfy (\ref{def23}).
The output is a defining polynomial, as in (\ref{def23}), for the extension of $\mQ$ containing the supplied numbers, along with an expression for each supplied number in terms of $x$ as in (\ref{expsalg}).

That we can do this is a powerful result in mathematics due to Artin known as the \textbf{primitive element theorem}:
If $E\supseteq F$ is a finite degree separable (\textit{i.e.}, the minimal polynomial has distinct roots) field extension, then $E=F[\alpha]$ for some $\alpha\in E$.
Given that we have a field extension $E=\mQ[\sqrt2, \sqrt3]$, we seek to express this as $E=\mQ[\alpha]$, where $\alpha$ is the primitive element.
The independence of $1$, $\sqrt2$, $\sqrt3$, and $\sqrt6$ over the rationals means that $\alpha$ cannot be generated by $\sqrt2$, $\sqrt3$, or $\sqrt6$ alone.
This exhausts all subfields of degree two.
Hence, it must be the entire field.
The primitive element is $x = \sqrt2 + \sqrt3$, and (\ref{def23}) tells us how to express $\sqrt2$ and $\sqrt3$ in terms of $x$.

\end{document}